\documentclass[journal]{IEEEtran}

\usepackage{cite}
\usepackage[dvips]{graphicx}
\usepackage[cmex10]{amsmath}
\usepackage{color}
\usepackage{bm}
\usepackage{theorem}

{\theorembodyfont{\normalfont}
\theoremheaderfont{\normalfont\it}
\newtheorem{thm}{ Theorem}
\newtheorem{dfn}[thm]{ Definition}
\newtheorem{lmm}[thm]{ Lemma}

\newtheorem{crl}[thm]{ Corollary}
}

{\theorembodyfont{\normalfont}
\theoremheaderfont{\normalfont\it}
}

{\theorembodyfont{\normalfont}
\theoremheaderfont{\normalfont\it}
}

{\theorembodyfont{\normalfont}
\theoremheaderfont{\normalfont\it}
}

\newcommand{\bra}[1]{\mbox{$\left\langle#1\right|$}}
\newcommand{\ket}[1]{\mbox{$\left|#1\right\rangle$}}
\newcommand{\inpro}[2]{\mbox{$\left\langle#1|#2\right\rangle$}}
\newcommand{\outpro}[2]{\mbox{$\ket{#1}\!\bra{#2}$}}
\newcommand{\proj}[1]{\mbox{$\ket{#1}\!\bra{#1}$}}

\onecolumn

\begin{document}

\title{Asymptotic Compressibility of \\Entanglement and Classical Communication in \\Distributed Quantum Computation}

\author{Eyuri~Wakakuwa and Mio Murao
\thanks{This work is supported in part by Project for Developing Innovation Systems of Ministry of Education, Culture, Sports, Science and Technology (MEXT), Japan. The work of E. Wakakuwa is supported by ALPS. The work of M. Murao is supported from JSPS by KAKENHI (Grant No. 23540463 and 26330006). This work was presented in part at the 2014 Quantum Information Processing workshop.}
\thanks{E. Wakakuwa is with the Department of Physics, Graduate School of Science, The University of Tokyo (email: wakakuwa@eve.phys.s.u-tokyo.ac.jp).}
\thanks{M. Murao is with the Department of Physics, Graduate School of Science, The University of Tokyo, and is with Institute for Nano Quantum Information Electronics, The University of Tokyo.}
}



\maketitle

\begin{abstract}
We consider implementations of a bipartite unitary on many pairs of unknown input states by local operation and classical communication assisted by shared entanglement. We investigate to what extent the entanglement cost and the classical communication cost can be compressed by allowing nonzero but vanishing error in the asymptotic limit of infinite pairs. We show that a lower bound on the minimal entanglement cost, the forward classical communication cost, and the backward classical communication cost per pair is given by the Schmidt strength of the unitary. We also prove that an upper bound on these three kinds of the cost is given by the amount of randomness that is required to partially decouple a tripartite quantum state associated with the unitary. In the proof, we construct a protocol in which quantum state merging is used. For generalized Clifford operators, we show that the lower bound and the upper bound coincide. We then apply our result to the problem of distributed compression of tripartite quantum states, and derive a lower and an upper bound on the optimal quantum communication rate required therein.
\end{abstract}


%
\IEEEpeerreviewmaketitle

\section{Introduction}
\label{sec:intro}

\IEEEPARstart{O}{ne} of the major interests in quantum information theory is to reveal the interconvertibility among different types of resources, such as quantum channels, noisy and noiseless entanglement, classical channels, and classical correlations. In the asymptotic limit of many copies, the rates at which those resources are interconvertible are derived in coding theorems. A theoretical framework composed of such coding theorems are called quantum Shannon theory \cite{deve04, deve08, min08}. Recently, it is shown that most of the central coding theorems in quantum Shannon theory can be systematically derived from protocols called quantum state merging \cite{horo05,horo07}, the fully quantum Slepian-Wolf \cite{abey09}, or its generalization called quantum state redistribution \cite{deve08_2,jon09,ming08,jonat08}. The main technical tool in their proof is the decoupling theorem \cite{patr08, fred10, fred10_2}. Main tasks considered in quantum Shannon theory such as {\it transmitting information} and {\it establishing entanglement} turned out to be convertible into the task of {\it destroying correlation between a system and the reference system}.

In classical information theory, however, not only {\it transmitting information} but also {\it computing a function} is an object of study \cite{ma11, gamal11}. Suppose that two distant parties Alice and Bob are given random bits ${\bm x}=(x^{(1)},\cdots,x^{(l)})$ and ${\bm y}=(y^{(1)},\cdots,y^{(l)})$, respectively, and try to compute a function $f({\bm x},{\bm y})$. Such a task is called distributed function computation. One way to compute $f$ is that Alice sends all ${\bm x}$ to Bob and Bob locally computes $f({\bm x},{\bm y})$, which apparently requires $l$ bits of communication. But depending on $f$, there are cases when the communication cost can be reduced below $l$ bits. For example, if $l=2$ and $f({\bm x},{\bm y})=x^{(1)}\cdot y^{(1)}+y^{(2)}$, only 1 bit of communication is sufficient, {because only Alice have to send $x^{(1)}$ to Bob}. Moreover, when Alice and Bob are given ${\bm x}_1,\cdots,{\bm x}_n$ and ${\bm y}_1,\cdots,{\bm y}_n$ from i.i.d. information sources, and try to compute $f({\bm x}_1,{\bm y}_1),\cdots,f({\bm x}_n,{\bm y}_n)$, there are cases when the classical communication cost can be {\it compressed} depending on $f$. That is, by allowing nonzero but vanishing error in the limit of $n\rightarrow\infty$, the total communication cost per input pair can be reduced below the single-shot limit.

\IEEEpubidadjcol

A quantum analog of distributed function computation would be distributed quantum computation \cite{cirac99, yim04}. Among many tasks, one of the simplest is implementations of bipartite unitaries on unknown input states by local operation and classical information (LOCC) assisted by shared entanglement. Suppose Alice and Bob have quantum systems $A$ and $B$ in unknown states $|\psi\rangle^A$ and $|\phi\rangle^B$, respectively, and try to apply a unitary $U^{AB}$ by entanglement-assisted LOCC (EALOCC). A trivial way is one using quantum teleportation \cite{bennett93}, where Alice teleports her input system to Bob, Bob performs the unitary, and then he teleports Alice's input system back to her. In the case of two-qubit unitaries, such a protocol consumes two Bell pairs and two bits of classical communication in both direction. But in \cite{eisert00}, a protocol is proposed for performing two-qubit controlled-unitary gates deterministically and exactly by using one ebit of entanglement resource and one bit of classical communication in both direction. This result indicates that, as well as the classical distributed function computation, there are cases where we can reduce the cost of resources depending on the unitary. 

Many studies have been made on such a task, particularly to find more efficient protocols which consume less resources \cite{eisert00, cirac01, groisman05, chen05, ye06, berry07, zhao08, yu10, cohen10}, and to derive minimum amount of resources that are required to accomplish the task \cite{soeda11, stahlke11}. However, finding out efficient protocols is difficult in general. Moreover, most studies have only focused on single-shot protocols. 

In the present paper, we investigate compressibility of the entanglement cost and the classical communication cost in EALOCC implementations of bipartite unitaries by considering an asymptotic scenario. We consider a task of applying $(U^{AB})^{\otimes n}$ on $(\ket{\Phi_d}^{AR_A}\ket{\Phi_d}^{BR_B})^{\otimes n}$ by EALOCC, where $R_A$ and $R_B$ are inaccessible reference systems, ${\rm dim}{\mathcal H}^A={\rm dim}{\mathcal H}^B=d$ and $\Phi_d$ is a $d$-dimensional maximally entangled state. { T}his is equivalent to performing $(U^{AB})^{\otimes n}$ on an arbitrary unknown input state $\ket{\psi}\in({\mathcal H}^A\otimes{\mathcal H}^B)^{\otimes n}$. We allow nonzero but vanishing error in the asymptotic limit of $n\rightarrow\infty$. We investigate the minimum amount of the entanglement cost, the forward and the backward classical communication cost per input pair in the asymptotic limit. 

We prove that a lower bound on the minimum amount of the three costs {per pair} is given by a parameter called the Schmidt strength of the unitary ({\it Theorem \ref{thm:lowerbound}}). To derive an upper bound, we focus on the fact that applying $(U^{AB})^{\otimes n}$ on $(|\Phi_{d}^{AR_A}\rangle|\Phi_{d}^{BR_B}\rangle)^{\otimes n}$ is equivalent to transforming $|\Psi(U^{\dagger})\rangle^{\otimes n}:=({U^{\dagger}}^{AB}|\Phi_{d}^{AR_A}\rangle|\Phi_{d}^{BR_B}\rangle)^{\otimes n}$ into $(|\Phi_{d}^{AR_A}\rangle|\Phi_{d}^{BR_B}\rangle)^{\otimes n}$. {That is, for a given state $|\Psi(U^{\dagger})\rangle^{\otimes n}$, Alice and Bob need to decouple $AR_A$ and $BR_B$, while preserving entanglement between $AB$ and $R_AR_B$.} For this task, we construct a three-turn protocol consisting of two steps. In the first step, Alice performs a measurement on her system, sends the result to Bob, and Bob applies a unitary on his system. In the second step, Bob sends a part of his system to Alice by quantum state merging. We find that Alice's measurement must decouple $AR_A$ and $R_B$ ({\it Lemma \ref{lmm:alicedecouple}}). That is, we need to find a quantum operation on $A$ that decouples $AR_A$ and $R_B$. The cost of randomness required in this decoupling process turns out to be equal to the entanglement cost, the forward and the backward {classical} communication cost in the protocol ({\it Theorem \ref{thm:upperbound}}). Thus we derive an upper bound on the optimal amount of the three costs in terms of the decoupling cost.

For generalized Clifford operators, we prove that the lower bound and the upper bound coincide, thus we derive the optimal entanglement cost, the forward and the backward classical communication cost ({\it Theorem \ref{thm:clifcoin}}).

We then relate our results to an apparently different task in quantum Shannon theory, namely, distributed compression of multipartite quantum states. In distributed compression, multiple senders $A_1,\cdots, A_m$ initially share $n$ identical copies of a pure state $|\psi\rangle^{A_1\dots A_mR}$, where $R$ is a reference system. The task is to compress their shares and transmit them to a receiver with a vanishingly small error, only by quantum communication from each of the $m$ parties to the receiver. Contrary to the bipartite setting \cite{ahn06,abey09}, little has been known on more than bipartite case since it was first formulated and analyzed in \cite{avis08}. In the present paper, we consider distributed compression of tripartite states associated with bipartite unitaries in a specific setting, and derive a lower and an upper bound on the optimal quantum communication rate required therein.

The structure of this paper is as follows. {In Section \ref{sec:preliminaries}, we review results from previous studies on EALOCC implementations of bipartite unitaries in the single-shot regime.} In Section \ref{sec:formulation} we give the formulation of our problem. In Section \ref{sec:lowerbound}, we derive a lower bound on the three kinds of the cost. In Section \ref{sec:singleshot}, we consider single-shot protocols for implementing bipartite unitaries by three-turn LOCC started by Alice's measurement. We derive conditions on Alice's measurement. In Section \ref{sec:upperbound}, using the result in Section \ref{sec:singleshot}, we derive an upper bound on the three kinds of the cost. In Section \ref{sec:clifford}, we prove that for generalized Clifford operators, the lower bound and the upper bound coincide. In Section \ref{sec:distributedcompression}, we apply our results to the problem of distributed compression of tripartite quantum states. Conclusions and discussions are given in Section \ref{sec:conclusion}.

{\it Notations.} We abbreviate $(M^A\otimes I^B)\ket{\psi}^{AB}$ as $M^A\ket{\psi}^{AB}$ and $(M^A\otimes I^B)\rho^{AB}(M^A\otimes I^B)^{\dagger}$ as $M^A\rho^{AB}M^{A\dagger}$. For $\ket{\phi}^{AB}$, $\phi^{A}$ {and ${\rm Tr}_B[\ket{\phi}]$} represents ${\rm Tr}_B[\proj{\phi}]$. {When ${\mathcal E}^A$ is a quantum operation on $A$, we abbreviate $({\mathcal E}^A\circ{\rm id}^B)(\rho^{AB})$ as ${\mathcal E}^A(\rho^{AB})$. ${\mathcal E}(|\phi\rangle)$ represents ${\mathcal E}(\proj{\phi})$.} We denote a system composed of $n$ identical systems $A$ as $A^n$ and $\bar{A}$. The fidelity of the states $\rho$ and $\sigma$ is defined as $F(\rho,\sigma)=({\rm Tr}\sqrt{\sqrt{\rho}\sigma\sqrt{\rho}})^2$. {We abbreviate $F(\rho,\proj{\phi})$ as $F(\rho,\ket{\phi})$.}

\section{Preliminaries}
\label{sec:preliminaries}
In EALOCC implementations of bipartite unitaries, Alice and Bob apply a bipartite unitary $U^{AB}$ on unknown quantum states $\varphi^A\in{\mathcal H}^A$ and $\psi^B\in{\mathcal H}^B$, by LOCC using some resource entanglement shared between the parties in advance. Protocols for this task are classified in terms of the success probability and the fidelity of the final state to the target state $U^{AB}\ket{\varphi}^A\ket{\psi}^B$. A protocol is called {\it deterministic} if it succeeds in implementing $U^{AB}$ with the probability one, otherwise it is called {\it probabilistic}. A protocol is called {\it exact} if the fidelity of the final state to the target state is one, otherwise it is called {\it approximate}.

Many studies have been made to find more efficient protocols which consume less resources \cite{eisert00, cirac01, groisman05, chen05, ye06, berry07, zhao08, yu10, cohen10}. A probabilistic and approximate protocol for two-qubit unitaries is proposed and investigated in \cite{cirac01}. Probabilistic and exact protocols for two-qubit unitaries are studied in \cite{groisman05, chen05, ye06}. A deterministic and exact protocol for two-qubit unitaries is proposed in \cite{eisert00}, and those for general bipartite unitaries are studied in \cite{yu10, cohen10}. In \cite{soeda11}, the minimum entanglement cost of deterministic and exact protocols for two-qubit controlled-unitaries is investigated. It is shown that any protocol for that task requires at least 1 ebit of entanglement when the resource state is a pure entanglement with the Schmidt rank 2, despite the fact that the controlled-unitary can be almost equal to the identity operator. The result is generalized for arbitrary bipartite unitaries in \cite{stahlke11}. It is also shown numerically in \cite{stahlke11} that there exists a class of two-qubit controlled-unitaries which can be implemented exactly and deterministically by using entanglement resource with the Schmidt rank 3, but with the entanglement entropy less than 1 ebit.

A difficulty in investigating EALOCC implementations of bipartite unitaries originates from the fact that we have no guiding principle in searching efficient protocols. Consequently, previous approaches are mostly case-by-case studies which cannot be straightforwardly generalized. In what follows, we show that the decoupling point of view, which is a powerful tool in quantum Shannon theory, may be one of such guiding principles. Although our main focus is an asymptotic scenario, our decoupling-based method would also be helpful in investigating single-shot protocols, as it is indicated in Section \ref{sec:singleshot}.

\section{Formulation of the problem}
\label{sec:formulation}

\begin{figure}[t]
\begin{center}
\includegraphics[bb={0 0 549 203}, scale=0.445]{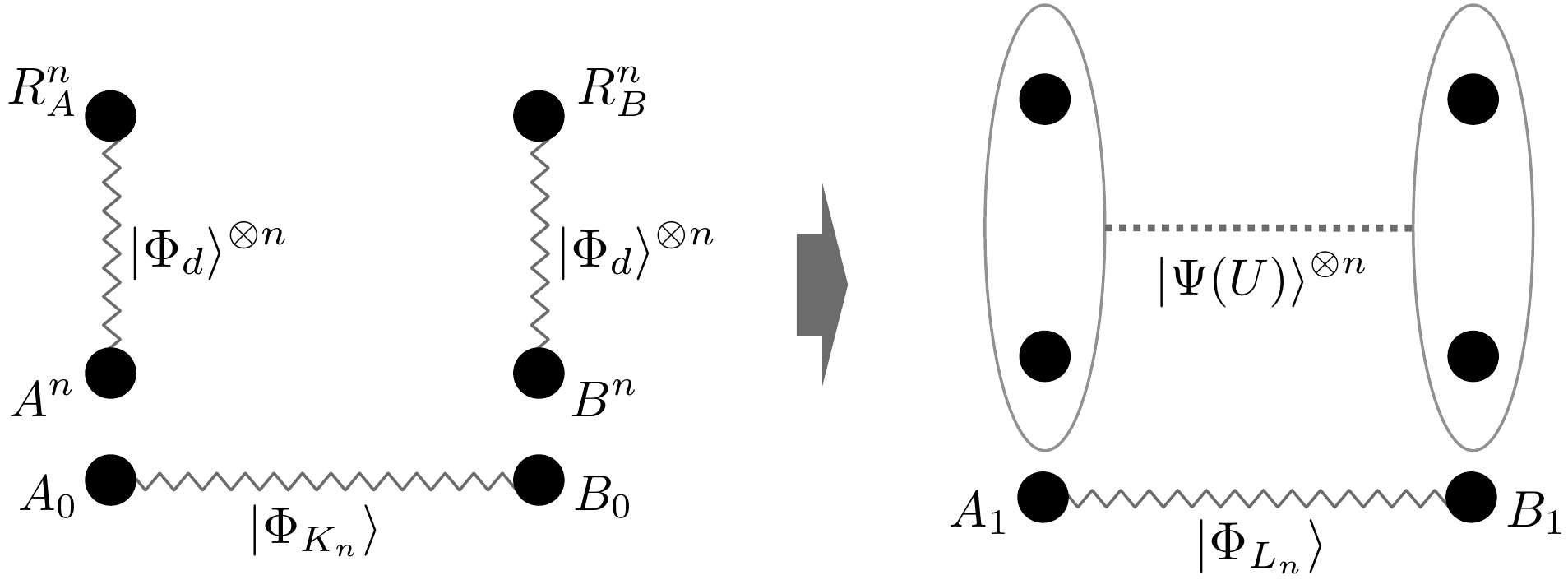}
\end{center}
\caption{{ The task we consider in this paper is to apply $(U^{AB})^{\otimes n}$ on $(\ket{\Phi_d}^{AR_A}\ket{\Phi_d}^{BR_B})^{\otimes n}$ by using the resource entanglement $\Phi_{K_n}^{A_0B_0}$. The fact that $R_A$ and $R_B$ are inaccessible to Alice and Bob makes the problem nontrivial. The entanglement cost is defined as the difference between the initial entanglement and the final entanglement shared by Alice and Bob.}}
\label{fig:dfntask}
\end{figure}

{We consider a task in which Alice and Bob apply a bipartite unitary $U^{AB}$ on unknown quantum states $\varphi^A\in{\mathcal H}^A$ and $\psi^B\in{\mathcal H}^B$, by LOCC using a resource state $\phi^{A_0B_0}$ shared in advance.} We assume that ${\rm dim}{\mathcal H}^A={\rm dim}{\mathcal H}^B=d$. The equivalent task is that Alice and Bob { apply $U^{AB}$} on $|\Phi_d\rangle^{AR_A}|\Phi_d\rangle^{BR_B}$ by LOCC with a resource state $\phi^{A_0B_0}$. Here, $\Phi_d$ is a $d$-dimensional maximally entangled state. $R_A$ and $R_B$ are reference systems that are inaccessible and invisible to Alice and Bob.

As an asymptotic version of the protocol, we consider a task in which Alice and Bob applies $(U^{AB})^{\otimes n}$ on $(|\Phi_d\rangle^{AR_A}|\Phi_d\rangle^{BR_B})^{\otimes n}$ by LOCC using a resource state $\Phi_{K_n}^{A_0B_0}$ (Fig. \ref{fig:dfntask}). Again $R_A$ and $R_B$ are reference systems that Alice and Bob cannot access. We evaluate the efficiency of the protocol by the fidelity between the final state of the protocol and the desired state $(U^{AB}\ket{\Phi_d}^{AR_A}\ket{\Phi_d}^{BR_B})^{\otimes n}$. We do not require that the fidelity is unity for finite $n$. Instead, we require that the fidelity converges to one in the limit of $n\rightarrow\infty$. A rigorous definition is below.

\begin{dfn}
Consider a unitary $U:{\mathcal H}^A\otimes{\mathcal H}^B\rightarrow{\mathcal H}^A\otimes{\mathcal H}^B$ acting on two $d$-level systems $A$ and $B$. { Let $|\Psi(U)\rangle:=U^{AB}\ket{\Phi_d}^{AR_A}\ket{\Phi_d}^{BR_B}$.} Let Alice and Bob have registers $A_0$, $A_1$ and $B_0$, $B_1$, respectively. We refer to the following quantum operation ${\mathcal M}_n$ as an EALOCC implementation of $U^{\otimes n}$ with the error $\epsilon_n$, the entanglement cost $\log{K_n}-\log{L_n}$, the forward classical communication cost $C_n^{\rightarrow}$, and the backward classical communication cost $C_n^{\leftarrow}$. Here, ${\mathcal M}_n:A^nA_0\otimes B^nB_0\rightarrow A^nA_1\otimes B^nB_1$ is a LOCC and
\begin{eqnarray}
F(\rho({\mathcal M_n}), |\Psi(U)\rangle^{\otimes n}|\Phi_{L_n}\rangle^{A_1B_1})\geq1-\epsilon_n
\label{eq:fidelityn}
\end{eqnarray}
for $\rho({\mathcal M}_n)={\mathcal M}_n(|\Phi_{d}^{AR_A}\rangle^{\otimes n}|\Phi_{d}^{BR_B}\rangle^{\otimes n}|\Phi_{K_n}\rangle^{A_0B_0})$. $C_n^{\rightarrow}$ and $C_n^{\leftarrow}$ is  the total amount of classical communication transmitted from Alice to Bob and Bob to Alice, respectively, in ${\mathcal M}_n$, measured by bits. 
\end{dfn}
{
\begin{dfn}
A rate triplet $(R_E,C^{\rightarrow},C^{\leftarrow})$ is said to be achievable if there exists a sequence of EALOCC implementations of $U^{\otimes n}$ such that $\epsilon_n\rightarrow0$, $\frac{1}{n}(\log{K_n}-\log{L_n})\rightarrow R_E$, $\frac{1}{n}C_n^{\rightarrow}\rightarrow C^{\rightarrow}$ and $\frac{1}{n}C_n^{\leftarrow}\rightarrow C^{\leftarrow}$ in the limit of $n\rightarrow\infty$. The set of all achievable rate triplets is called the rate region.
\end{dfn}

Let ${\hat{\mathcal M}}_n$ be the action of ${\mathcal M}_n$ on $A^nB^n$. The condition (\ref{eq:fidelityn}) implies that, for {\it almost all} input states $|\phi\rangle\in{\mathcal H}^{A^n}\otimes{\mathcal H}^{B^n}$, the final state ${\hat{\mathcal M}}_n(\phi^{A^nB^n})$ is sufficiently close to the desired state $U^{\otimes n}|\phi\rangle$. Indeed, due to the relation between entanglement fidelity and ensemble fidelity \cite{schumacher96}, the condition (\ref{eq:fidelityn}) implies
\begin{eqnarray}
\int_{\phi} p(d\phi)\:F({\hat{\mathcal M}}_n(\phi),U^{\otimes n}|\phi\rangle)\geq1-\epsilon_n,
\label{eq:ensfide}
\end{eqnarray}
where the average is taken with respect to the Haar measure on ${\mathcal H}^{A^n}\otimes{\mathcal H}^{B^n}$ (see Appendix \ref{app:avfid}). On the other hand, we do not require that the protocol is {\it universal} in the sense that ${\hat{\mathcal M}}_n(\phi^{A^nB^n})$ is close to the desired state $U^{\otimes n}|\phi\rangle$ for {\it all} input states $|\phi\rangle\in{\mathcal H}^{A^n}\otimes{\mathcal H}^{B^n}$.}

{ In the form of the resource inequality \cite{deve04, deve08}, we are trying to find the set of rate triplets $(R_E,C^{\rightarrow},C^{\leftarrow})$ that satisfies
\begin{eqnarray}
R_E[qq]+C^{\rightarrow}[c\rightarrow c]+C^{\leftarrow}[c\leftarrow c]\geq\left\langle U^{AB}:\frac{1}{d^2}I^{AB}\right\rangle.\nonumber\\
\label{eq:resineq}
\end{eqnarray}
The R.H.S. is the {\it relative resource} in the sense of \cite{devetak06}, which represents an operation on $AB$ that is guaranteed to behave like the unitary $U$ only if the reduced average state of the input is close to $(\frac{1}{d^2}I^{AB})^{\otimes n}$. 
}

\section{Lower Bound}
\label{sec:lowerbound}

Any bipartite unitary acting on ${\mathcal H}_A\otimes{\mathcal H}_B$ is decomposed as
\begin{eqnarray}
U^{AB}=\sum_{s=1}^Sc_s{E}^{A}_s\otimes{F}^{B}_s,\nonumber
\end{eqnarray}
where $c_s>0, {E}_s\in{\mathcal L}({\mathcal H}^A),{ F}_s\in{\mathcal L}({\mathcal H}^B), \sum_{k=1}^Sc_s^2=1$ and $d^{-1}{\rm Tr}[{ E}_s^{\dagger}{ E}_{s'}]=d^{-1}{\rm Tr}[{ F}_s^{\dagger}{ F}_{s'}]=\delta_{ss'}$. { It is called the operator Schmidt decomposition and $S$ is called the operator Schmidt number.} The Schmidt strength of the unitary $U$ is defined as $K(U):=-\sum_{s=1}^Sc_s^2\log{c_s^2}$ \cite{niel03,oppenheim04}. Let $\ket{\phi_s}:=E_s^A\ket{\Phi_d}^{AR_A}$ and $\ket{\psi_s}:=F_s^A\ket{\Phi_d}^{BR_B}$. Then we have $\inpro{\psi_s}{\psi_s'}=\inpro{\phi_s}{\phi_s'}=\delta_{ss'}$ and $\ket{\Psi(U)}=\sum_{s=1}^Sc_s\ket{\psi_s}^{AR_A}\ket{\phi_s}^{BR_B}$. Thus the { entanglement entropy} of $\ket{\Psi(U)}$ between $AR_A$ and $BR_B$ is equal to $K(U)$.

\begin{thm}
A rate triplet $(R_E,C^{\rightarrow},C^{\leftarrow})$ is achievable only if $R_E,C^{\rightarrow},C^{\leftarrow}\geq K(U)$.
\label{thm:lowerbound}
\end{thm}

\begin{IEEEproof}
$R_E\geq K(U)$ follows from the monotonicity of entanglement under LOCC. To prove $C^{\rightarrow},C^{\leftarrow}\geq K(U)$, we show that if Alice and Bob can apply $U^{AB}$ on $\ket{\Phi_d}^{AR_A}\ket{\Phi_d}^{BR_B}$, Alice can communicate $K(U)$ bits of classical information to Bob. Because of the symmetry, Bob can also communicate $K(U)$ bits to Alice. Since the classical communication power cannot exceed the classical communication cost, we have $C^{\rightarrow},C^{\leftarrow}\geq K(U)$. 

The proof is based on the protocol proposed in \cite{dominic07}. Suppose that Alice and Bob initially share $\ket{\Phi_d}^{AB_2}\ket{\Phi_d}^{BB_1}$. Let $\{\sigma_i^A\}_{i=1}^{d^2}$ be the set of unitaries on $A$ satisfying $d^{-1}{\rm Tr}[{\sigma_i^{\dagger}\sigma_j}]=\delta_{ij}$. (For example, we can take the set of generalized Pauli operators on $A$.) Alice sends the message $i$ with the probability $p_i=1/d^2$ by performing $\sigma_i^A$ and then applying $U^{AB}$. The state after these two operations is
\begin{eqnarray}
\ket{\Psi_i}&=&U^{AB}\sigma_i^A\ket{\Phi_d}^{AB_2}\ket{\Phi_d}^{BB_1}\nonumber\\
&=&U^{AB}(\sigma_i^T)^{B_2}\ket{\Phi_d}^{AB_2}\ket{\Phi_d}^{BB_1}.\nonumber
\end{eqnarray}
The entropy of the reduced state on Bob's systems is
\begin{eqnarray}
S(BB_1B_2)_{\Psi_i}=S(A)_{\Psi_i}=\log{d}.\nonumber
\end{eqnarray}
On the other hand, the average state {on Bob's side} is
\begin{eqnarray}
\bar{\Psi}^{BB_1B_2}&=&\frac{1}{d^2}\sum_{i=1}^{d^2}{\rm Tr}_{A}[U^{AB}(\sigma_i^T)^{B_2}\ket{\Phi_d}^{AB_2}\ket{\Phi_d}^{BB_1}]\nonumber\\
&=&\frac{1}{d^2}I^{B_2}\otimes{\rm Tr}_A[U^{AB}(I^{A}\otimes{\Phi}^{BB_1})U^{\dagger AB}].\nonumber
\end{eqnarray}
Thus the entropy of the state is
\begin{eqnarray}
S(BB_1B_2)_{\bar \Psi}=S(B_2)_{\bar \Psi}+S(BB_1)_{\bar \Psi}=\log{d}+K(U).\nonumber
\end{eqnarray}
The Holevo information \cite{schumacher97,holevo98} is then given by
\begin{eqnarray}
\chi(\{p_i,\Psi_i^{BB_1B_2}\})&=&S(BB_1B_2)_{\bar \Psi}-\frac{1}{d^2}\sum_{i=1}^{d^2}S(BB_1B_2)_{\Psi_i}\nonumber\\
			&=&K(U).\nonumber
\end{eqnarray}
Hence Alice can communicate $K(U)$ bits of classical information to Bob per use of $U^{AB}$ on $\ket{\Phi_d}^{AB_2}\ket{\Phi_d}^{BB_1}$.
\end{IEEEproof}

\section{Single-shot protocols}
\label{sec:singleshot}

To derive an upper bound on the optimal {rate} of the entanglement cost and the classical communication costs in both direction, namely, to prove that a certain rate triplet $(R_E,C^{\rightarrow},C^{\leftarrow})$ is achievable, we need to prove the existence of a protocol by which we can achieve the rate triplet. In this paper, we consider three-turn protocols in which Alice first preforms a measurement on her system, sends the result to Bob, Bob performs a measurement on his system, sends the result to Alice, and then Alice performs an operation on her system. Without loss of generality, we assume that Alice's last operation is to apply an isometry. 

In this section, we consider single-shot protocols for applying $U^{AB}$ on $|\Phi_d\rangle^{AR_A}|\Phi_d\rangle^{BR_B}$ by three-turn EALOCC. We derive conditions on Alice's first measurement for the protocol to succeed in high fidelity. The results obtained here are then applied to the asymptotic case in the next section.

\begin{figure}[t]
\begin{center}
\includegraphics[bb={0 0 548 404}, scale=0.45]{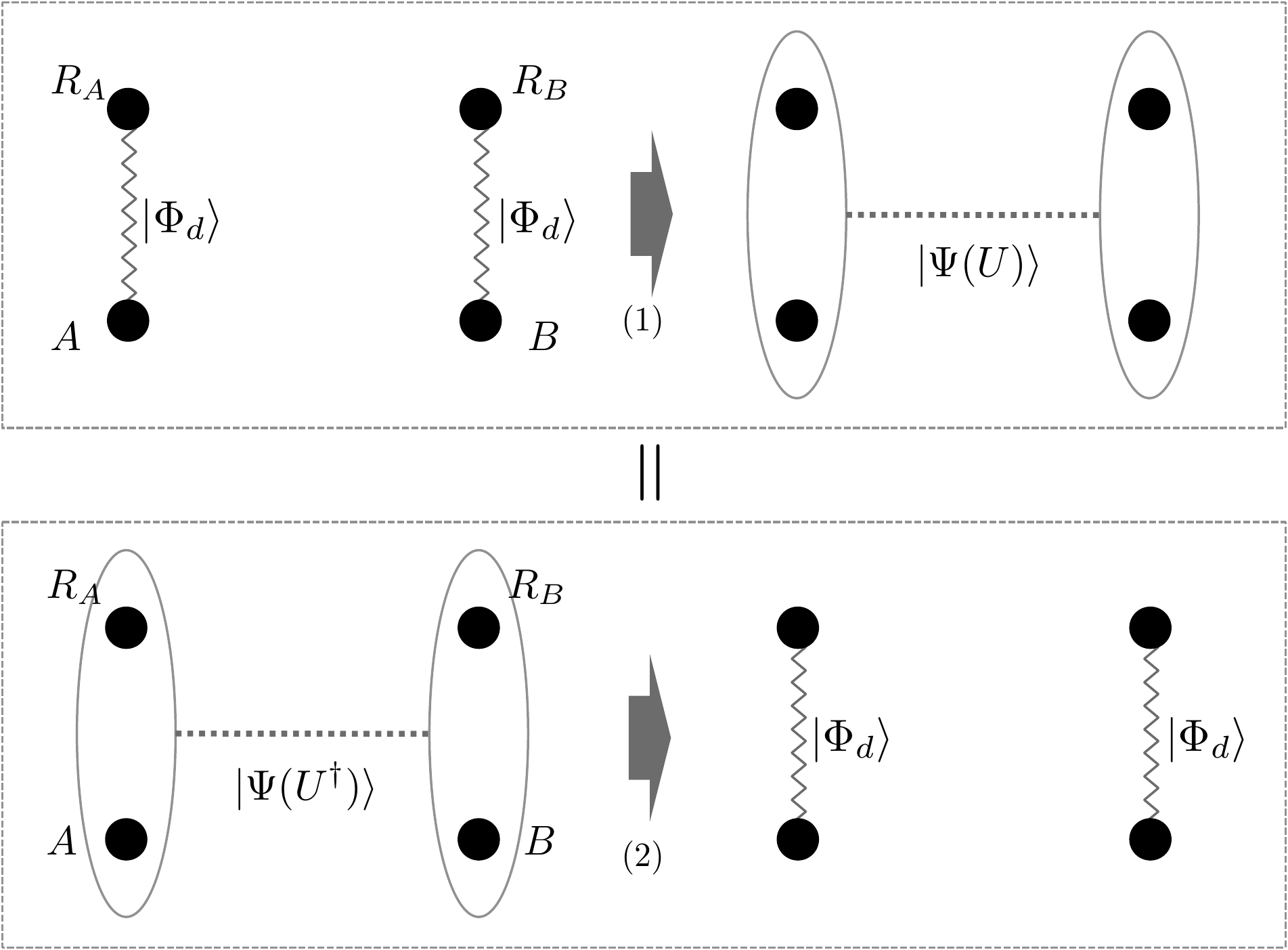}
\end{center}\caption{The task of performing $U^{AB}$ on $|\Phi_d\rangle^{AR_A}|\Phi_d\rangle^{BR_B}$ {represented by the process}  (1) is equivalent to transforming $|\Psi(U^{\dagger})\rangle$ into $|\Phi_d\rangle^{AR_A}|\Phi_d\rangle^{BR_B}$ {represented by the process}  (2). {This equivalencies due to the fact that} $R_A$ and $R_B$ are reference systems that are inaccessible and invisible to Alice and Bob.}
\label{fig:equivalenttask}
\end{figure}

Since $R_A$ and $R_B$ are reference systems that Alice and Bob cannot access, we can take any pure state on $ABR_AR_B$ for the initial state, as long as {it is a $d$-dimensional maximally entangled state between $AB$ and $R_AR_B$}. Indeed, if a quantum operation ${\mathcal E}$ on $AB$ satisfies ${\mathcal E}^{AB}(|\Phi_d\rangle^{AR_A}|\Phi_d\rangle^{BR_B})=U^{AB}|\Phi_d\rangle^{AR_A}|\Phi_d\rangle^{BR_B}$, it also satisfies ${\mathcal E}^{AB}(U_0^{AB}|\Phi_d\rangle^{AR_A}|\Phi_d\rangle^{BR_B})=(UU_0)^{AB}|\Phi_d\rangle^{AR_A}|\Phi_d\rangle^{BR_B}$ for any $U_0$. In particular, we can take $U_0$ to be $U^{\dagger}$. Thus the task of applying $U^{AB}$ on $|\Phi_d\rangle^{AR_A}|\Phi_d\rangle^{BR_B}$ is equivalent to transforming $|\Psi(U^{\dagger})\rangle^{ABR_AR_B}:={U^{\dagger}}^{AB}|\Phi_{d}^{AR_A}\rangle|\Phi_{d}^{BR_B}\rangle$ into $|\Phi_d\rangle^{AR_A}|\Phi_d\rangle^{BR_B}$ (Fig. \ref{fig:equivalenttask}). Observe that both $|\Psi(U^{\dagger})\rangle$ and $|\Phi_d\rangle|\Phi_d\rangle$ are $d^2$-dimensional maximally entangled state between $AB$ and $R_AR_B$. While $|\Phi_d\rangle|\Phi_d\rangle$ obviously has no correlation between $AR_A$ and $BR_B$, $|\Psi(U^{\dagger})\rangle$ has some amount of entanglement depending on $U^{\dagger}$. Thus, for a given initial state $\ket{\Psi(U^{\dagger})}^{ABR_AR_B}$, Alice and Bob need to decouple $AR_A$ and $BR_B$ while preserving the maximal entanglement between $AB$ and $R_AR_B$.

Consider a three-turn EALOCC protocols for transforming $|\Psi(U^{\dagger})\rangle^{ABR_AR_B}$ into $|\Phi_d\rangle^{AR_A}|\Phi_d\rangle^{BR_B}$ by using the resource state $\ket{\phi_{\rm res}}^{A_0B_0}$. Since Alice's last operation does not change the reduced state of $BR_BR_A$, $|\Phi_d\rangle^{BR_B}$ must be obtained after Bob's operation. In order that this is possible, Alice's first measurement should be such that $AR_A$ and $R_B$ is decoupled for all measurement outcomes. Conversely, if Alice's measurement satisfies this condition, Bob can obtain $|\Phi_d\rangle^{BR_B}$ by performing an isometry on his system depending on her measurement  {outcome}. To be precise, we have the following lemmas.

\begin{lmm}
\label{lmm:alicedecouple}
Let ${\mathcal M}:AA_0\otimes BB_0\rightarrow AA_1\otimes BB_1$ be an LOCC such that
\begin{eqnarray}
F(\rho({\mathcal M}),|\Psi(U)\rangle)\geq1-\epsilon,
\label{eq:1shoterror}
\end{eqnarray}
where $\rho({\mathcal M}):={\rm Tr}_{A_1B_1}[{\mathcal M}(|\Phi_d\rangle^{AR_A}|\Phi_d\rangle^{BR_B}|\phi_{\rm res}\rangle^{A_0B_0})]$ and $\phi_{\rm res}$ is a pure resource state shared in advance. Suppose ${\mathcal M}$ is a three-turn protocol that starts with Alice's measurement described by measurement operators $\{{ M}_k^{AA_0\rightarrow A'}\}_{k=1}^K$. Let $p_k=\|{ M}_k^{AA_0\rightarrow A'}|\Psi(U^{\dagger})\rangle|\phi_{\rm res}\rangle\|_1^2$ and $|\Psi_k(U^{\dagger})\rangle=p_k^{-1/2}{ M}_k^{AA_0\rightarrow A'}|\Psi(U^{\dagger})\rangle|\phi_{\rm res}\rangle$. Then
\begin{eqnarray}
\sum_{k=1}^Kp_k\left\|\Psi_k(U^{\dagger})^{A'R_AR_B}-\Psi_k(U^{\dagger})^{A'R_A}\otimes\frac{1}{d}{ I}^{R_B}\right\|_1\leq4\sqrt{\epsilon}.\nonumber\\
\label{eq:1shotdistance}
\end{eqnarray}
\end{lmm}

\begin{IEEEproof}
The condition (\ref{eq:1shoterror}) is equivalent to
\begin{eqnarray}
F(\rho({\mathcal M},U^{\dagger})^{ABR_AR_B},\ket{\Phi_d}^{AR_A}\ket{\Phi_d}^{BR_B})\geq1-\epsilon,\nonumber
\end{eqnarray}
where $\rho({\mathcal M},U^{\dagger}):={\rm Tr}_{A_1B_1}[{\mathcal M}(|\Psi(U^{\dagger})\rangle^{AR_ABR_B}|\phi_{\rm res}\rangle^{A_0B_0})]$. Let $\{{ N}_{l}^{(k)BB_0\rightarrow BB_1}\}_{l=1}^L$ be measurement operators of Bob's measurement. We assume that Alice's second operation is an isometry ${ V}^{(kl)A'\rightarrow AA_1}$.  Define $p_{kl}=\|({ M}_k\otimes{ N}_{l}^{(k)})|\Psi(U^{\dagger})\rangle|\phi_{\rm res}\rangle\|_1^2$, $|\Psi_{kl}(U^{\dagger})\rangle=p_{kl}^{-1/2}({ M}_k\otimes{ N}_{l}^{(k)})|\Psi(U^{\dagger})\rangle|\phi_{\rm res}\rangle$ and $|\Psi'_{kl}(U^{\dagger})\rangle={ V}^{(kl)}|\Psi_{kl}(U^{\dagger})\rangle$. From (\ref{eq:1shoterror}), we have
\begin{eqnarray}
1-\epsilon&\leq&\sum_{kl}p_{kl}F(\Psi'_{kl}(U^{\dagger})^{ABR_AR_B},\ket{\Phi_d}^{AR_A}\ket{\Phi_d}^{BR_B})\nonumber\\
&=&\sum_{kl}p_{kl}F(\ket{\Psi'_{kl}(U^{\dagger})},\ket{\Phi_d}^{AR_A}\ket{\Phi_d}^{BR_B}\ket{\phi_{kl}}^{A_1B_1})\nonumber
\end{eqnarray}
for some states $\phi_{kl}$. Using the relation $\|\rho-\sigma\|_1\leq2\sqrt{1-F(\rho,\sigma)}$ and the convexity of the trace distance, we obtain
\begin{eqnarray}
	2\sqrt{\epsilon}&\geq&\sum_{kl}p_{kl}\left\|\Psi'_{kl}(U^{\dagger})-\Phi_d^{AR_A}\otimes\Phi_d^{BR_B}\otimes\phi_{kl}^{A_1B_1}\right\|_1\nonumber\\
	&=&\sum_{kl}p_{kl}\left\|\Psi_{kl}(U^{\dagger})^{A'R_ABB_0R_B}-V^{(kl)}(\Phi_d^{AR_A}\otimes\Phi_d^{BR_B}\otimes\phi_{kl}^{A_1B_1})V^{(kl)\dagger}\right\|_1\nonumber\\
	&\geq&\sum_{kl}p_{kl}\left\|\Psi_{kl}(U^{\dagger})^{A'R_AR_B}-V^{(kl)}(\Phi_d^{AR_A}\otimes\phi_{kl}^{A_1})V^{(kl)\dagger}\otimes\frac{1}{d}I^{R_B}\right\|_1\nonumber\\
	&\geq&\sum_{k}p_{k}\left\|\sum_{l}p_{l|k}\Psi_{kl}(U^{\dagger})^{A'R_AR_B}-\psi_{k}^{A'R_A}\otimes\frac{1}{d}I^{R_B}\right\|_1.\nonumber
\end{eqnarray}
Here we defined $\psi_{k}:=\sum_{l}p_{l|k}V^{(kl)}(\Phi^{AR_A}\otimes\phi_{kl}^{A_1})V^{(kl)\dagger}$. Observe that
\begin{eqnarray}
\sum_{l}p_{l|k}\Psi_{kl}(U^{\dagger})=\frac{1}{p_k}\sum_{l}({ M}_k\otimes{ N}_l^{(k)})(\Psi(U^{\dagger})\otimes\phi_{\rm res})({ M}_k\otimes{ N}_l^{(k)})^{\dagger}\nonumber
\end{eqnarray}
and, since ${\mathcal N}:\rho\rightarrow\sum_{l=1}^L{ N}_l^{(k)}\rho{ N}_l^{(k)\dagger}$ is a CPTP map on $BB_0$, 
\begin{eqnarray}
\sum_{l}p_{l|k}\Psi_{kl}(U^{\dagger})^{A'R_AR_B}=\Psi_{k}(U^{\dagger})^{A'R_AR_B}.\nonumber
\end{eqnarray}
Hence we obtain
\begin{eqnarray}
2\sqrt{\epsilon}&\geq&\sum_{k}p_{k}\left\|\Psi_{k}(U^{\dagger})^{A'R_AR_B}-\psi_{k}^{A'R_A}\otimes\frac{1}{d}I^{R_B}\right\|_1\nonumber\\
&\geq&\sum_{k}p_{k}\left\|\Psi_{k}(U^{\dagger})^{A'R_A}-\psi_{k}^{A'R_A}\right\|_1\nonumber\\
&=&\sum_{k}p_{k}\left\|\Psi_{k}(U^{\dagger})^{A'R_A}\otimes\frac{1}{d}I^{R_B}-\psi_{k}^{A'R_A}\otimes\frac{1}{d}I^{R_B}\right\|_1\nonumber
\end{eqnarray}
By the triangle inequality, we obtain
\begin{eqnarray}
	4\sqrt{\epsilon}&\geq&\sum_{k}p_{k}\left\|\Psi_{k}(U^{\dagger})^{A'R_AR_B}-\psi_{k}^{A'R_A}\otimes\frac{1}{d}I^{R_B}\right\|_1+\sum_{k}p_{k}\left\|\Psi_{k}(U^{\dagger})^{A'R_A}\otimes\frac{1}{d}I^{R_B}-\psi_{k}^{A'R_A}\otimes\frac{1}{d}I^{R_B}\right\|_1\nonumber\\
	&\geq&\sum_{k=1}^Kp_k\left\|\Psi_k(U^{\dagger})^{A'R_AR_B}-\Psi_k(U^{\dagger})^{A'R_A}\otimes\frac{1}{d}{ I}^{R_B}\right\|_1.\nonumber
\end{eqnarray}
\end{IEEEproof}

\begin{lmm}
If (\ref{eq:1shotdistance}) is satisfied and $|\Psi^p_k(U^{\dagger})\rangle^{A'R_AB_0}$ is a purification of $\Psi_k(U^{\dagger})^{A'R_A}$, there exists a set of unitaries $\{{ W}^{(k)}\}_{k=1}^K$ on $BB_0$ such that 
\begin{eqnarray}
&&\sum_{k=1}^Kp_k\left\|{ W}^{(k)}\Psi_k(U^{\dagger})^{A'R_AB_0BR_B}{ W}^{(k)\dagger}-\Psi^p_k(U^{\dagger})^{A'R_AB_0}\otimes\Phi_d^{BR_B}\right\|_1\leq4\sqrt[4]{\epsilon}.\nonumber
\label{eq:1shoterror2}
\end{eqnarray}
\label{lmm:uhlmann}
\end{lmm}
\begin{IEEEproof}
{ It is straightforward from Uhlmann's theorem \cite{uhlmann76,jozsa94}. See Lemma 2.2 in \cite{deve08} and Proposition 3 in \cite{horo07}.}
\end{IEEEproof}

\begin{figure}[t]
\begin{center} \includegraphics[bb={0 0 716 309}, scale=0.34]{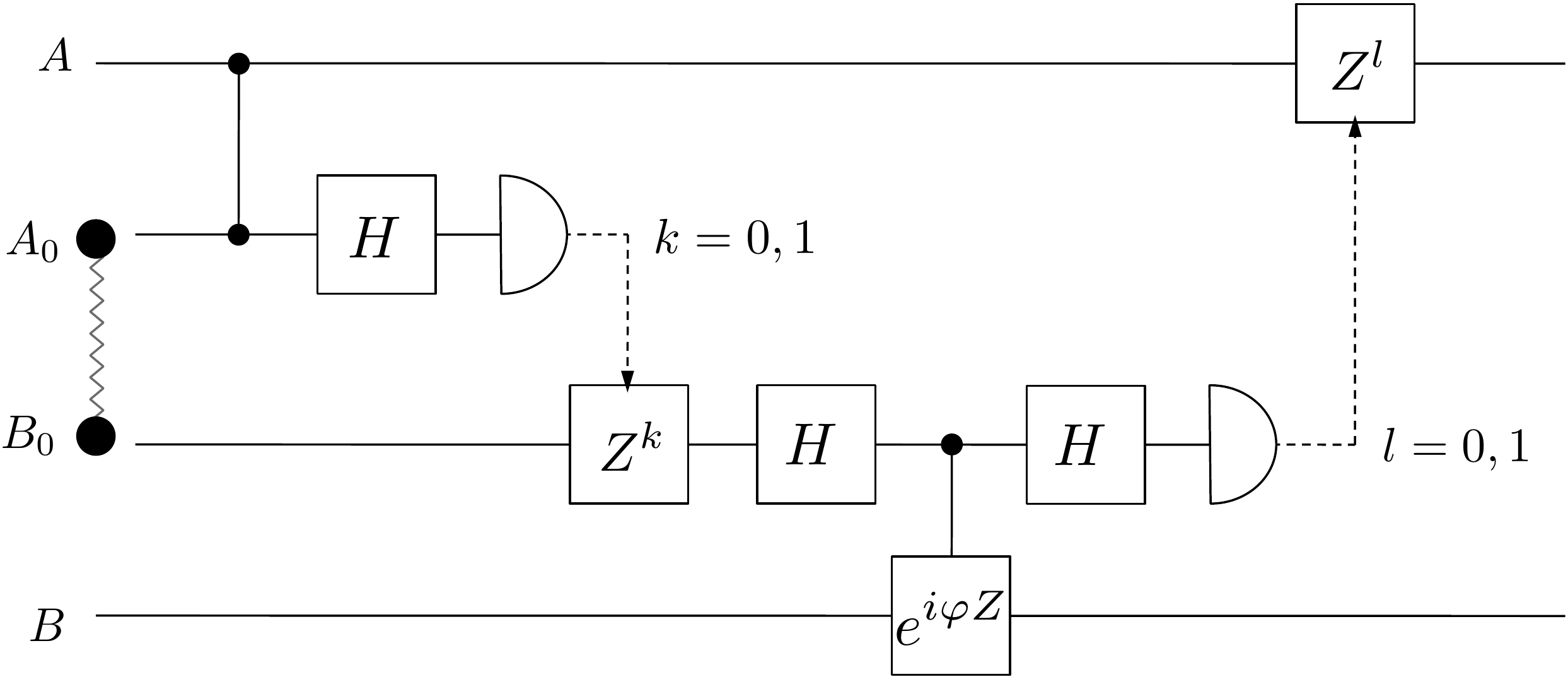}
\end{center}\caption{{ The protocol for implementing controlled-phase gate proposed in \cite{eisert00}. $H$ denotes the Hadamard gate and the measurements are performed with the computational basis.}}
\label{fig:cphase}
\end{figure}

{ {\it Example.} To illustrate the implications of Lemma \ref{lmm:alicedecouple} and Lemma \ref{lmm:uhlmann}, we consider a protocol for performing two-qubit controlled-phase gate exactly and deterministically by EALOCC, which is proposed in \cite{eisert00}. Let $U=\proj{0}\otimes I+\proj{1}\otimes e^{i\varphi Z}$. The protocol for implementing this unitary is equivalent to the protocol depicted in Fig. \ref{fig:cphase}. In this case $\ket{\phi_{\rm res}}=(\ket{00}+\ket{11})/\sqrt{2}$, $A'=A$, $p_k=1/2\:(k=1,2)$ and $M_k^{AA_0\rightarrow A}=\bra{0}^{A_0}\otimes I^A\pm\bra{1}^{A_0}\otimes Z^{A}$, where $+$ and $-$ corresponds to $k=1$ and $2$, respectively. We have
\begin{eqnarray}
\ket{\Psi(U^{\dagger})}&=&\frac{1}{2}\left(\ket{00}+e^{+i\varphi}\ket{11}\right)^{AR_A}\ket{00}^{BR_B}+\frac{1}{2}\left(\ket{00}+e^{-i\varphi}\ket{11}\right)^{AR_A}\ket{11}^{BR_B}\nonumber\\
				&=:&\frac{1}{\sqrt{2}}\ket{\Phi_{+\varphi}}^{AR_A}\ket{00}^{BR_B}+\frac{1}{\sqrt{2}}\ket{\Phi_{-\varphi}}^{AR_A}\ket{11}^{BR_B}\nonumber
\end{eqnarray}
and, in particular, we have
\begin{eqnarray}
\Psi(U^{\dagger})^{AR_AR_B}=\frac{1}{{2}}\Phi_{+\varphi}^{AR_A}\otimes\proj{0}^{R_B}+\frac{1}{{2}}{\Phi_{-\varphi}}^{AR_A}\otimes\proj{1}^{R_B}.\nonumber
\end{eqnarray}
By using $\phi_{\rm res}^{A_0}=I/2$, we obtain
\begin{eqnarray}
\Psi_k(U^{\dagger})^{AR_AR_B}&=&p_k^{-1}{\rm Tr}_{BB_0}[M_k|\Psi(U^{\dagger})\rangle|\phi_{\rm res}\rangle]\nonumber\\
&=&p_k^{-1}M_k\left(\Psi(U^{\dagger})^{AR_AR_B}\otimes\frac{1}{2}I^{A_0}\right)M_k^{\dagger}\nonumber\\
&=&{\mathcal T}_{\rm deph}^{A}(\Psi(U^{\dagger})^{AR_AR_B}),\nonumber
\end{eqnarray}
where ${\mathcal T}_{\rm deph}^{A}$ is the dephasing channel on $A$ defined as
\begin{eqnarray}
{\mathcal T}_{\rm deph}^{A}(\rho)=\frac{1}{2}(\rho+Z\rho Z).\nonumber
\end{eqnarray}
Since
\begin{eqnarray}
{\mathcal T}_{\rm deph}^{A}(\Phi_{+\varphi}^{AR_A})={\mathcal T}_{\rm deph}^{A}(\Phi_{-\varphi}^{AR_A})=\frac{1}{2}(\proj{00}+\proj{11}),\nonumber
\end{eqnarray}
we have
\begin{eqnarray}
{\mathcal T}_{\rm deph}^{A}(\Psi(U^{\dagger})^{AR_AR_B})=\frac{1}{2}(\proj{00}+\proj{11})^{AR_A}\otimes\frac{1}{2}I^{R_B}.\nonumber
\end{eqnarray}
Thus $\Psi_k(U^{\dagger})\;(k=1,2)$ satisfies the condition (\ref{eq:1shotdistance}) with $\epsilon=0$.

Alice's measurement operation described by $M_k$ with $\phi_{\rm res}$ is equivalent to performing the controlled-Z gate on $B_0A$ with $B_0$ initially prepared in the state $\ket{+}=(\ket{0}+\ket{1})/\sqrt{2}$, up to phases on $B_0$. Indeed, we have
\begin{eqnarray}
M_k\ket{\phi_{\rm res}}=\ket{0}^{B_0}\otimes I^A\pm\ket{1}^{B_0}\otimes Z^A\;(k=1,2).\nonumber
\end{eqnarray}
The one bit classical communication from Alice to Bob and the following $Z$ gate on $B_0$ eliminate the phase that depends on the measurement outcome. Bob then applies a unitary on $B_0B$ to obtain $|\Psi^p(U^{\dagger})\rangle^{AR_AB_0}|\Phi_{2}\rangle^{BR_B}$, where $|\Psi^p(U^{\dagger})\rangle$ is a purification of $\Psi_k(U^{\dagger})^{AR_A}$. In the remaining part of the protocol, Alice and Bob obtain $\Phi_{2}^{AR_A}$ from $|\Psi^p(U^{\dagger})\rangle^{AR_AB_0}$.

}

\begin{figure*}[t]
\begin{center}
\includegraphics[bb={0 0 848 200}, scale=0.55]{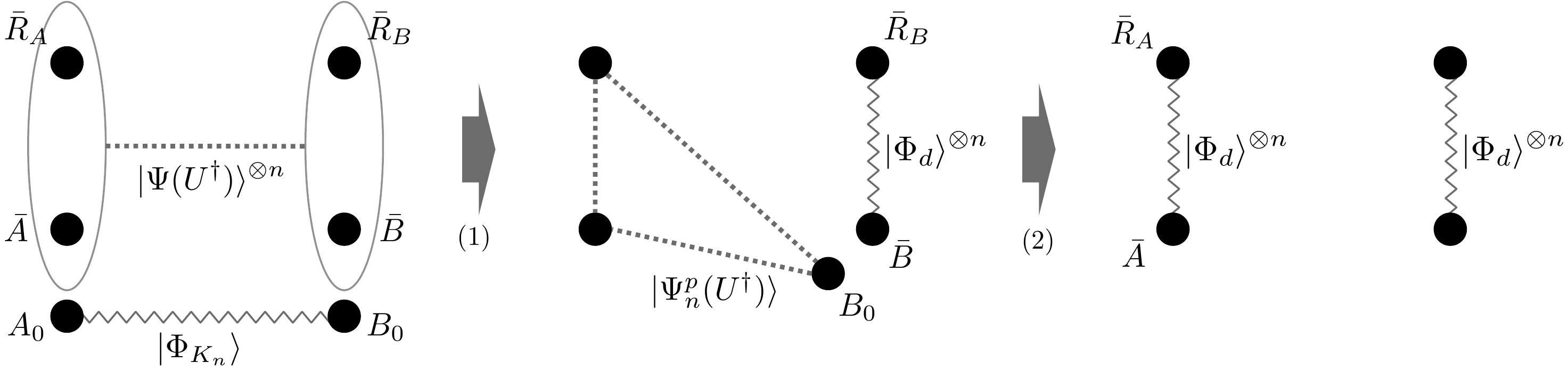}
\end{center}\caption{We consider {a} three-turn EALOCC protocol for implementing $U^{\otimes n}$ on $|\Phi_d^{\otimes n}\rangle^{\bar{A}\bar{R}_A}|\Phi_d^{\otimes n}\rangle^{\bar{B}\bar{R}_B}$. (1) In the `forward process', Alice and Bob recover $|\Phi_d^{\otimes n}\rangle^{\bar{B}\bar{R}_B}$ from $|\Psi(U^{\dagger})^{\otimes n}\rangle^{\bar{A}\bar{B}\bar{R}_A\bar{R_B}}$ by using a resource state $\ket{\Phi_{K_n}}^{A_0B_0}$. (2) After that, they recover $|\Phi_d^{\otimes n}\rangle^{\bar{A}\bar{R}_A}$ by {performing} quantum state merging from Bob to Alice. The entanglement cost in {step (2)} is negligible in the asymptotic limit.}
\label{fig:protocol}
\end{figure*}

\section{Upper Bound}
\label{sec:upperbound}

Let us return to the asymptotic case. {We consider a protocol where $L_n=1$, that is, no entanglement is left after the protocol}. {We denote $A^n$ as $\bar{A}$ and $B^n$ as $\bar{B}$, and so on.} The task is to apply a unitary $(U^{AB})^{\otimes n}$ on $(|(\Phi_d\rangle^{AR_A}|\Phi_d\rangle^{BR_B})^{\otimes n}$ by using $|\Phi_{K_n}\rangle^{A_0B_0}$ as a resource, or, equivalently, to transform $\ket{\Psi(U^{\dagger})^{\otimes n}}^{\bar{A}\bar{B}\bar{R}_A\bar{R}_B}\ket{\Phi_{K_n}}^{A_0B_0}$ into $|\Phi_d^{\otimes n}\rangle^{\bar{A}\bar{R}_A}|\Phi_d^{\otimes n}\rangle^{\bar{B}\bar{R}_B}$. For this task, we construct a protocol consisting of two {processes} (Fig. \ref{fig:protocol}). In the `forward process' aiming at obtaining $|\Phi_{d}^{BR_B}\rangle^{\otimes n}$ from $|\Psi(U^{\dagger})\rangle^{\otimes n}|\Phi_{K_n}\rangle^{A_0B_0}$, Alice performs a measurement on her system, sends the result to Bob, and Bob applies a unitary on his system. It is followed by the `backward process' consisting of quantum state merging from Bob to Alice, aiming at obtaining $|\Phi_{d}^{AR_A}\rangle^{\otimes n}$. From the result in the previous section, Alice's measurement must be such that ${\bar A}{\bar R}_A$ and ${\bar R}_B$ is almost decoupled for all measurement outcomes. Also, if the decoupling succeeds in high fidelity, Bob can obtain $|\Phi_{d}^{BR_B}\rangle^{\otimes n}$ by applying a unitary on his system depending on Alice's measurement outcome. 

\begin{figure}[t]
\begin{center} \includegraphics[bb={0 0 492 306}, scale=0.36]{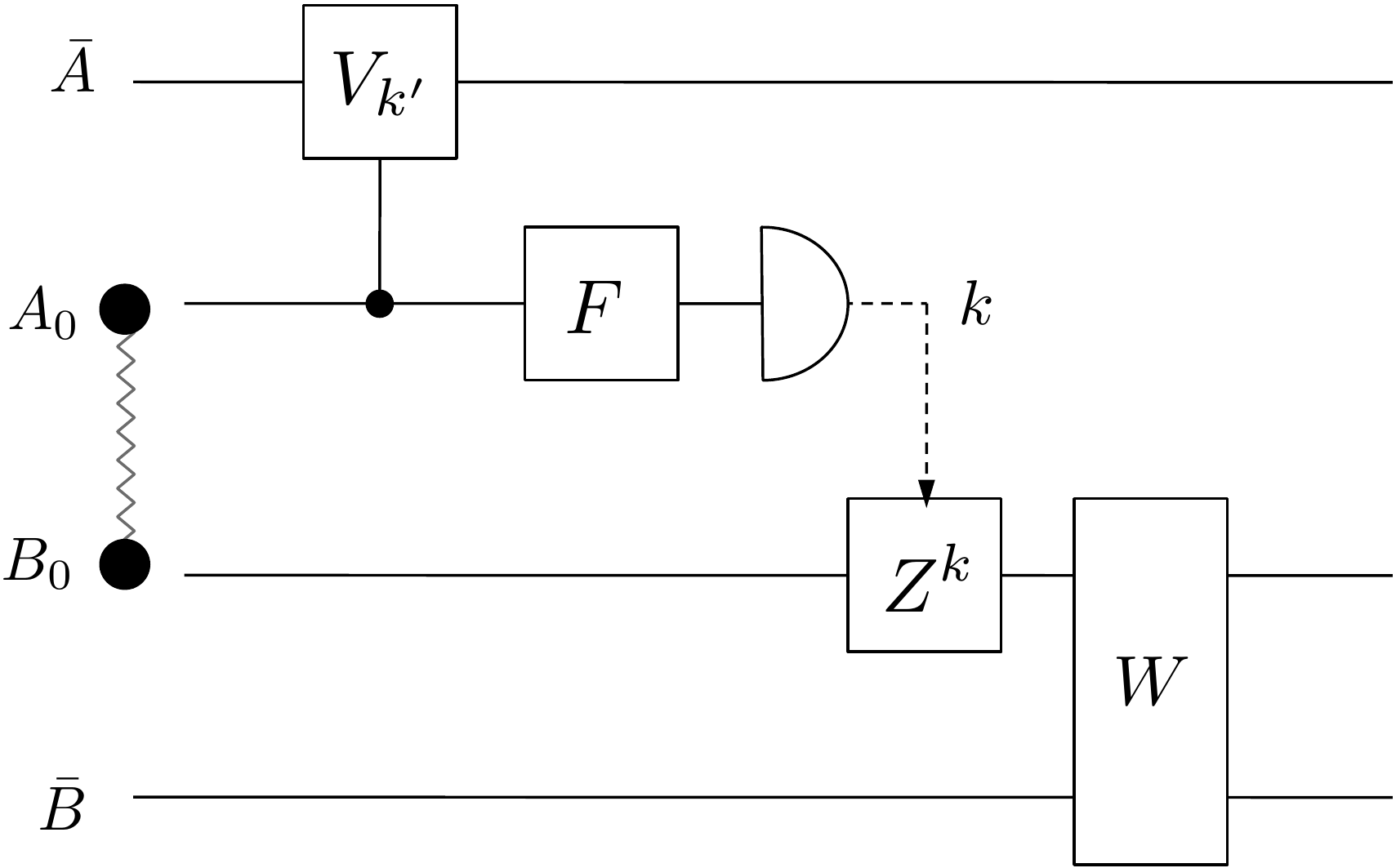}
\end{center}\caption{{ As the `forward process', we consider a protocol similar to the first half of the protocol presented in Fig. \ref{fig:cphase}. $F$ denotes the Fourier transform and $Z$ is the phase gate. The measurement is performed with the computational basis. This scheme is similar to the protocol proposed in \cite{cohen10}.}}
\label{fig:forwardp}
\end{figure}

As the forward process, we consider a protocol consisting of the following steps (Fig. \ref{fig:forwardp}).
\begin{enumerate}
\item Alice applies a controlled-unitary gate of the form ${\tilde V}^{\bar{A}A_0}=\sum_{k'=1}^{K_n}\proj{k'}^{A_0}\otimes V_{k'}^{\bar A}$.
\item Alice applies a Fourier transform ${ F}=\frac{1}{\sqrt{K_n}}\sum_{kk'}e^{2\pi ikk'/{K_n}}\outpro{k}{k'}$ on $A_0$.
\item Alice performs a projective measurement on $A_0$ with the basis $\{\ket{k}\}_{k=1}^{K_n}$.
\item Alice sends the measurement result $k$ to Bob. This requires $\log{K_n}$ bits of forward classical communication.
\item Bob applies a phase gate $Z^k$ on $B_0$. Here, $Z=\sum_{k'=1}^{K_n}e^{-2\pi ik'/{K_n}}\outpro{k'}{k'}$.
\item Bob applies a unitary $W^{{\bar B}B_0}$.
\end{enumerate}

\begin{figure}[t]
\begin{center} \includegraphics[bb={0 0 481 170}, scale=0.52]{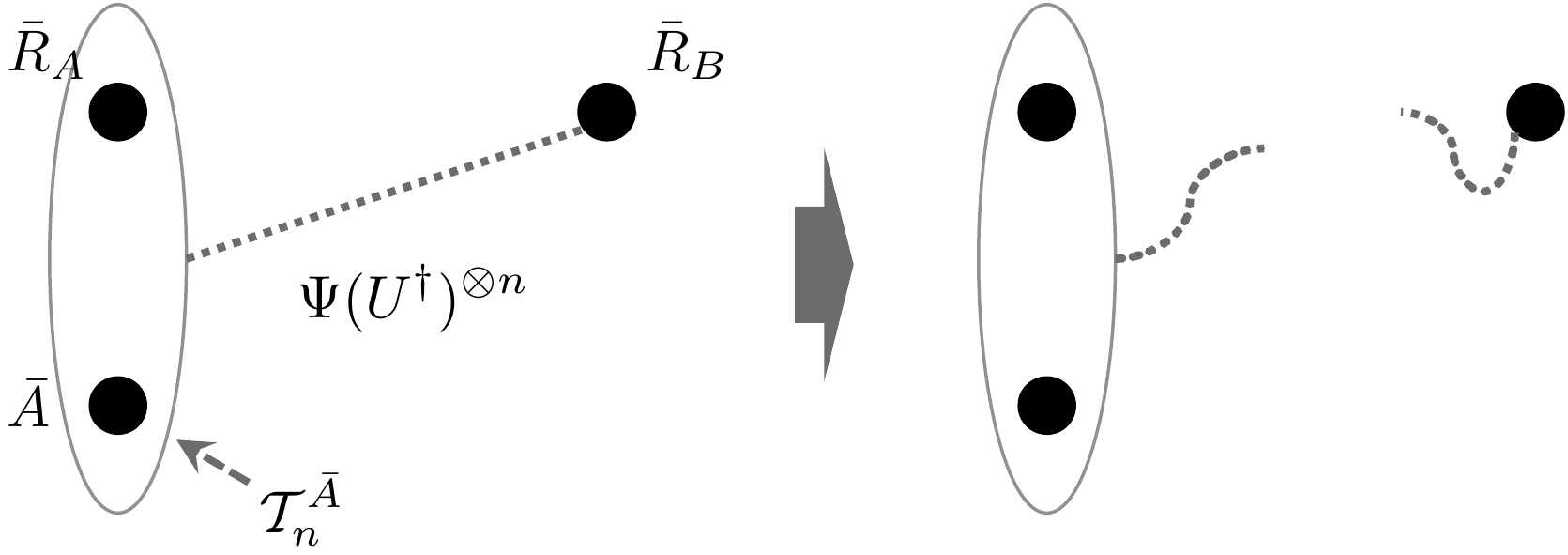}
\end{center}\caption{{ When we only consider ${\bar{A}\bar{R}_A\bar{R}_B}$, the action of the five steps in the forward process is equal to performing ${\mathcal T}_n^{\bar A}$ on $\bar{A}$. Here, ${\mathcal T}_n^{\bar A}$ is a random unitary operation defined as ${\mathcal T}_n^{\bar A}:\tau\mapsto2^{-nR}\sum_{k=1}^{2^{nR}}V_k\tau V_k^{\dagger}$. The correlation between $\bar{A}\bar{R}_A$ and $\bar{R}_B$ needs to be destroyed by this operation.}}
\label{fig:decoupling}
\end{figure}

The total action of the five steps is equivalent to applying a controlled-unitary gate ${\tilde V}^{{\bar A}B_0}=\sum_{k'=1}^{K_n}\proj{k'}^{B_0}\otimes V_{k'}^{\bar A}$ with $B_0$ initially prepared in the state $|+_{K_n}\rangle:=\frac{1}{\sqrt{K_n}}\sum_{k'=1}^{K_n}|k'\rangle$. Thus, regardless of the measurement outcome $k$, the state after step 5 is given by
\begin{eqnarray}
|{\Psi}_{n}^{5}(U^{\dagger})\rangle^{\bar{A}\bar{B}B_0\bar{R}_A\bar{R}_B}:=\frac{1}{\sqrt{K_n}}\sum_{k'=1}^{K_n}\ket{k'}^{B_0}\otimes V_{k'}^{\bar A}\ket{\Psi(U^{\dagger})^{\otimes n}}^{\bar{A}\bar{B}\bar{R}_A\bar{R}_B}.\nonumber
\label{eq:psi5}
\end{eqnarray}
In particular, its reduced state on $\bar{A}\bar{R}_A\bar{R}_B$ is given by
\begin{eqnarray}
{\Psi}_{n}^{5}(U^{\dagger})^{\bar{A}\bar{R}_A\bar{R}_B}=\frac{1}{K_n}\sum_{k'=1}^{K_n}V_{k'}^{{\bar A}}(\Psi(U^{\dagger})^{\otimes n})^{\bar{A}\bar{R}_A\bar{R}_B}V_{k'}^{\dagger\bar{A}}.\nonumber\\
\label{eq:randomunitary}
\end{eqnarray}
From Lemma \ref{lmm:uhlmann}, if
\begin{eqnarray}
{\Psi}_{n}^{5}(U^{\dagger})^{\bar{A}\bar{R}_A\bar{R}_B}\approx{\Psi}_{n}^{5}(U^{\dagger})^{\bar{A}\bar{R}_A}\otimes\frac{1}{d^n}I^{\bar{R}_B},
\label{eq:approxproduct}
\end{eqnarray}
and $|\Psi^p_{n}(U^{\dagger})\rangle^{\bar{A}B_0\bar{R}_A}$ is a purification of ${\Psi}_{n}^{5}(U^{\dagger})^{\bar{A}\bar{R}_A}$, there exists a unitary $W$ on ${\bar B}B_0$ such that  $W|\Psi_{n}^{5}(U^{\dagger})\rangle^{\bar{A}\bar{B}B_0\bar{R}_A\bar{R}_B}\approx|\Psi^p_{n}(U^{\dagger})\rangle^{\bar{A}B_0\bar{R}_A}\ket{\Phi_d^{\otimes n}}^{\bar{B}\bar{R}_B}$ (Fig. \ref{fig:decoupling}). Thus Bob can obtain $\ket{\Phi_d^{\otimes n}}^{\bar{B}\bar{R}_B}$. Both the entanglement cost and the classical communication in this process is equal to $\log{K_n}$. 

The remaining task is to recover $|\Phi_d^{\otimes n}\rangle^{\bar{A}\bar{R}_A}$ from $|\Psi^p_{n}(U^{\dagger})\rangle^{\bar{A}B_0\bar{R}_A}$. Here we introduce quantum state merging \cite{horo05,horo07} in which Bob merges $B_0$ to Alice. If this process succeeds in high fidelity, Alice can obtain $\ket{\Phi_{d}^{\otimes n}}^{\bar{A}\bar{R}_A}$ by performing local isometry on her system. In the asymptotic limit of many copies, {the entanglement cost per copy in quantum state merging is given by the quantum conditional entropy $S(B_0|\bar{A})_{\Psi^p_{n}(U^{\dagger})}$, and the classical communication cost per copy is equal to the quantum mutual information $I(B_0:{\bar R}_A)_{\Psi^p_{n}(U^{\dagger})}$}. {As we show in detail below, since $S(B_0|\bar{A})_{\Psi^p_{n}(U^{\dagger})}=0$, the entanglement cost in this process is negligible. Also, the classical communication cost is equal to $I(B_0:{\bar R}_A)_{\Psi^p_{n}(U^{\dagger})}=\log{K_n}$.} Thus the total amount of the entanglement cost, the forward and the backward communication cost per copy is $\frac{1}{n}\log{K_n}$. From (\ref{eq:randomunitary}) and (\ref{eq:approxproduct}), $K_n$ is equal to the cost of randomness required for decoupling $\bar{A}\bar{R}_A$ and $\bar{R}_B$ in the state $(\Psi(U^{\dagger})^{\otimes n})^{\bar{A}\bar{R}_A\bar{R}_B}$ by a random unitary operation on $\bar{A}$.

To be precise, we have the following statements.

\begin{dfn}
We say that $\Psi(U^{\dagger})^{AR_AR_B}$ is decoupled between $AR_A$ and $R_B$ with the randomness cost $R$ if, for any $\epsilon>0$ and for sufficiently large $n$, there exists a random unitary operation ${\mathcal T}_n^{\bar A}:\tau\mapsto2^{-nR}\sum_{k=1}^{2^{nR}}V_k\tau V_k^{\dagger}$ on $\bar{A}$ such that
\begin{eqnarray}
\left\|{\mathcal T}_n^{\bar A}(\Psi(U^{\dagger})^{\otimes n})^{\bar{A}\bar{R}_A\bar{R}_B}-{\mathcal T}_n^{\bar A}(\Psi(U^{\dagger})^{\otimes n})^{\bar{A}\bar{R}_A}\otimes \frac{1}{d^n}I^{\bar{R}_B}\right\|_1\leq\epsilon.
\label{eq:decouple1}
\end{eqnarray}
The partial decoupling cost of $U$ is defined as $D(U):=\inf\{R\:|\:\Psi(U^{\dagger})^{AR_AR_B}\text{ is decoupled between $AR_A$ and $R_B$}$ $\text{with the randomness cost }R\}$.
\label{dfn:randomcost}
\end{dfn}

\begin{thm}
A rate triplet $(R_E,C^{\rightarrow},C^{\leftarrow})$ is achievable if $R_E,C^{\rightarrow},C^{\leftarrow}\geq D(U)$.
\label{thm:upperbound}
\end{thm}

\begin{IEEEproof}
Take arbitrary $R>D(U)$ and let $K_n=2^{nR}$. Fix arbitrary $\delta>0$, choose sufficiently large $m$, let $\epsilon=\delta/{m^2}$ and choose sufficiently large $n$. Let ${\mathcal T}_n^{\bar A}:\tau\mapsto2^{-nR}\sum_{k=1}^{2^{nR}}V_k\tau V_k^{\dagger}$ be a random unitary operation that satisfies (\ref{eq:decouple1}). Using $V_k$, construct a controlled-unitary operator ${\tilde V}^{\bar{A}A_0}=\sum_{k'=1}^{K_n}\proj{k'}^{A_0}\otimes V_{k'}^{\bar A}$. Suppose that Alice and Bob performs the forward process. After step 5, the reduced state on $\bar{A}\bar{R}_A\bar{R}_B$, given by (\ref{eq:randomunitary}), is equal to ${\mathcal T}^{\bar A}_n(\Psi(U^{\dagger})^{\otimes n})^{\bar{A}\bar{R}_A\bar{R}_B}$. thus it satisfies
\begin{eqnarray}
\left\|\Psi_{n}^{5}(U^{\dagger})^{\bar{A}\bar{R}_A\bar{R}_B}-\Psi_{n}^{5}(U^{\dagger})^{\bar{A}\bar{R}_A}\otimes\frac{1}{d^n}I^{\bar{R}_B}\right\|_1\leq\epsilon.\nonumber
\label{eq:decouple2}
\end{eqnarray}
Hence, if $|\Psi^p_{n}(U^{\dagger})\rangle^{\bar{A}\bar{R}_AB_0}$ is a purification of ${\Psi}_{n}^{5}(U^{\dagger})^{\bar{A}\bar{R}_A}$, there exists a unitary $W$ on $\bar{B}B_0$ such that 
\begin{eqnarray}
\left\|{ W}\Psi_{n}^{5}(U^{\dagger}){ W}^{\dagger}-\Psi^p_{n}(U^{\dagger})^{\bar{A}\bar{R}_AB_0}\otimes(\Phi_{d}^{\otimes n})^{\bar{B}\bar{R}_B}\right\|_1\leq2\sqrt{\epsilon}.\nonumber
\label{eq:decouple3}
\end{eqnarray}
By applying $W$, Bob can obtain $|\Phi_d^{\otimes n}\rangle^{\bar{B}\bar{R}_B}$. 

Suppose that Alice and Bob perform this protocol on each of $m$ blocks of {length $n$ in} $\Psi(U^{\dagger})^{\otimes mn}$. Let $\Psi_{mn}^{5}(U^{\dagger}):=(\Psi_{n}^{5}(U^{\dagger}))^{\otimes m}$. In total, the state after step 5 satisfies
\begin{eqnarray}
\left\|\Psi_{mn}^{5}(U^{\dagger})-(\Psi^p_{n}(U^{\dagger})^{\otimes m})^{\bar{A}^m\bar{R}_A^mB_0^m}\otimes(\Phi_{d}^{\otimes mn})^{\bar{B}^m\bar{R}_B^m})\right\|_1\leq2m\sqrt{\epsilon}=2\sqrt{\delta}.
\label{eq:decouple4}
\end{eqnarray}

Now, consider quantum state merging of $|\Psi^p_{n}(U^{\dagger})\rangle^{\bar{A}\bar{R}_AB_0}$ from Bob to Alice. Since $S(B_0|\bar{A})=S(B_0\bar{A})-S(\bar{A})=S(\bar{R}_A)-S(\bar{A})=n\log{d}-n\log{d}=0$ for $\Psi^p_{n}(U^{\dagger})$, there exists a merging protocol of $\ket{\Psi^p_{n}(U^{\dagger})^{\otimes m}}^{\bar{A}^m\bar{R}_A^mB_0^m}$ with {error} $\delta$ and the entanglement cost $m\delta$. That is, there is an LOCC map ${\mathcal E}'$ that implements $|\Psi^p_{n}(U^{\dagger})^{\otimes m}\rangle^{\bar{A}^mB_0^m\bar{R}_A^m}|\Phi_{2^{m\delta}}\rangle^{A'_0B'_0}\rightarrow\ket{\Psi^p_{n}(U^{\dagger})^{\otimes m}}^{\bar{A}^m{\hat A}\bar{R}_A^m}$ with the error $\delta$. Here, $A'_0$, $B'_0$ and $\hat{A}$ are additional registers. Let ${O}:\bar{A}^m{\hat A}\rightarrow{\bar A}^m$ be the isometry such that ${ O}|\Psi^p_{n}(U^{\dagger})^{\otimes m}\rangle^{\bar{A}^m\bar{R}_A^m{\hat A}}=|\Phi_d^{\otimes mn}\rangle^{\bar{A}^m\bar{R}_A^m}$. Combining ${\mathcal E}'$ and $O$, we obtain an LOCC map ${\mathcal E}:\bar{A}^mB_0A'_0B'_0\rightarrow\bar{A}^m$ such that 
\begin{eqnarray}
\left\|{\mathcal E}\left(|\Psi^p_{n}(U^{\dagger})^{\otimes m}\rangle^{\bar{A}^m\bar{R}_A^mB_0}|\Phi_{2^{m\delta}}\rangle^{A'_0B'_0}\right)-(\Phi_{d}^{\otimes mn})^{\bar{A}^m\bar{R}_A^m}\right\|_1\leq{\delta}.
\label{eq:decouple5}
\end{eqnarray}
Combining (\ref{eq:decouple4}) and (\ref{eq:decouple5}), the total error is given by
\begin{eqnarray}
\left\|{\mathcal E}(|\Psi_{mn}^{5}(U^{\dagger})\rangle|\Phi_{2^{m\delta}}\rangle)-(\Phi_{d}^{\otimes mn})^{\bar{A}^m\bar{R}_A^m}\otimes(\Phi_{d}^{\otimes mn})^{\bar{B}^m\bar{R}_B^m}\right\|_1\leq2\sqrt{\delta}+\delta.\nonumber
\end{eqnarray}

In total, the entanglement cost in this protocol is $m\log{K_n}+m\delta=mn(R+\delta/n)$. The forward classical communication cost, which is given by the number of Alice's measurement outcomes, is $m\log{K_n}=mnR$. The backward classical communication cost is equal to $m(I(\bar{R}_A:B_0)_{\Psi_n^p(U^{\dagger})}+\delta)=m(S(B_0)_{\Psi_n^p(U^{\dagger})}+\delta)=m(\log{K_n}+\delta)=mn(R+\delta/n)$ bits. Thus the rate triplet $(R_E=R,C^{\rightarrow}=R,C^{\leftarrow}=R)$ is achievable, and consequently, any rate triplet $(R_E,C^{\rightarrow},C^{\leftarrow})$ is achievable if $R_E,C^{\rightarrow},C^{\leftarrow}\geq D(U)$.
\end{IEEEproof}

\begin{crl}
$D(U)\geq K(U)$.
\label{crl:kandd}
\end{crl}

\begin{IEEEproof}
Follows from Theorem \ref{thm:lowerbound} and Theorem \ref{thm:upperbound}. It can also be derived from Proposition 1 {presented} in \cite{berry08}.
\end{IEEEproof}
 
 {Obviously}, $\Psi(U^{\dagger})^{AR_AR_B}$ is decoupled between $AR_A$ and $R_B$ with the randomness cost $2\log{d}$, because applying uniformly random generalized Pauli operators on $\bar{A}$ decouples $\bar{A}$ and all the {other systems}. {Note that, since there is no correlation between $\bar{R}_A$ and $\bar{R}_B$ initially, decoupling $\bar{A}$ and $\bar{R}_A\bar{R}_B$  is sufficient for decoupling $\bar{A}\bar{R}_A$ and $\bar{R}_B$.} Thus we have $D(U)\leq2\log{d}$ and $(R_E,C^{\rightarrow},C^{\leftarrow})$ is achievable if $R_E,C^{\rightarrow},C^{\leftarrow}\geq2\log{d}$. However, this bound is trivial {since} the resource cost $R_E,C^{\rightarrow},C^{\leftarrow}=2\log{d}$ is achievable through the protocol in which Alice teleports her system to Bob, Bob applies the unitary, and teleports Alice's system back to her.

In analogy with the decoupling theorem for bipartite quantum states \cite{patr08, fred10, fred10_2,berry08}, we might expect that the random coding method with respect to the Haar distributed unitary ensembles on $\bar{A}$ is sufficient to derive $D(U)$. However, this is not the case. If we apply the Haar distributed random unitary on $\bar{A}$, not only the correlation between $\bar{A}\bar{R}_A$ and $\bar{R}_B$ but also the correlation between $\bar{A}$ and $\bar{R}_A$ is destroyed. Thus part of the total randomness is wastefully consumed for decoupling $\bar{A}$ and $\bar{R}_A$. In particular, if $K(U)$ is much small and the correlation between ${\bar A}$ and ${\bar R_A}$ is strong, much large amount of the randomness is wastefully consumed. To reduce the randomness cost, we need to find a random unitary operation on $\bar{A}$ which selectively decouples $\bar{A}\bar{R}_A$ and $\bar{R}_B$ without much destroying the correlation between $\bar{A}$ and $\bar{R}_A$.

We have not found how to calculate $D(U)$ in general, except for a particular class of bipartite unitaries which is discussed in the next section.

\section{Generalized Clifford Operators}
\label{sec:clifford}

In this section, we prove that $K(U_{\rm Cl})=D(U_{\rm Cl})$ for generalized Clifford operators $U_{\rm Cl}$, and thus the optimal entanglement cost, the forward and the backward classical communication cost per input pair is equal to $K(U_{\rm Cl})$. 

The generalized Pauli operators on $d$-dimensional Hilbert space is defined as $\sigma_{pq}:=X^pZ^q$, where $X:=\sum_{t=1}^{d}\outpro{t-1}{t}$ and $Z:=\sum_{t=1}^{d}e^{2\pi it/d}\outpro{t}{t}$ with a fixed basis $\{\ket{t}\}_{t=1}^d$. Here, subtraction is taken with mod $d$. A unitary $U$ on two $d$-dimensional Hilbert space ${\mathcal H}_A\otimes{\mathcal H}_B$ is called a generalized Clifford operator if, for any $p$, $q$, $r$ and $s$, there exist $p'$, $q'$, $r'$ and $s'$ such that $U(\sigma_{pq}\otimes\sigma_{rs})U^{\dagger}=\sigma_{p'q'}\otimes\sigma_{r's'}$.

The idea of the proof is as follows. Although $R_B$ is a reference system that Alice and Bob cannot access, for the moment imagine that we can apply a random unitary operation on $R_B$. In particular, suppose that we can randomly apply generalized Pauli operators $\sigma^T_{pq}$ with { the} uniform distribution. Because of {Schur's} lemma, the random Pauli operation decouples $AR_A$ and $R_B$. In the asymptotic limit of many copies of $\Psi(U^{\dagger})$, the number of Pauli operators {required} to decouple $AR_A$ and $R_B$ is equal to $I(AR_A:R_B)_{\Psi(U^\dagger)}=K(U)$ per copy. But {due to} the commutation relation of generalized Pauli and Clifford operators, we can replace $\sigma^T_{pq}$ on $R_B$ by $\sigma_{p'q'}\otimes\sigma_{r's'}$ on $AB$. In particular, for the state $\Psi(U^\dagger)^{AR_AR_B}$, applying $\sigma^T_{pq}$ on $R_B$ is equivalent to applying $\sigma_{p'q'}$ on $A$. Thus the random Pauli operation on $R_B$ is exactly substituted by a random Pauli operation on $A$. Hence the number of random Pauli operators on $A$ that we need to decouple $AR_A$ and $R_B$ is at most $I(AR_A:R_B)_{\Psi(U^\dagger)}=K(U)$ per copy. We show the rigorous proof below.

\begin{lmm}
(Proposition 2 in \cite{berry08}) 
For any $\rho^{AB}\in{\mathcal S}({\mathcal H}^A\otimes{\mathcal H}^B)$ and $\epsilon>0$, there {exists} $n_0$ such that for all $n\geq n_0$ the following holds. Let ${\mathcal H}^{A_{typ}}_{n,\epsilon}$ be the $\epsilon$-typical subspace of $(\rho^A)^{\otimes n}$ and let ${\Pi}^{A_{typ}}_{n,\epsilon}$ be the projection onto ${\mathcal H}^{A_{typ}}_{n,\epsilon}$.
Also let $\{p(dV),V\}$ be an ensemble of unitaries on ${\mathcal H}^{A_{typ}}_{n,\epsilon}$ such that for all $\ket{\phi}\in{\mathcal H}^{A_{typ}}_{n,\epsilon}$,
\begin{eqnarray}
\int_{V}p(dV)V\proj{\phi}V^{\dagger}=\frac{{\Pi}^{A_{typ}}_{n,\epsilon}}{{\rm Tr}{\Pi}^{A_{typ}}_{n,\epsilon}}.\nonumber
\end{eqnarray}
Then, if $R\geq I(A:B)_{\rho}+\epsilon$, there exists a set of unitaries $\{V_k\}^{2^{nR}}_{k=1}$ on the support of $p(dV)$ such that
\begin{eqnarray}
\left\|\frac{1}{2^{nR}}\sum_{k=1}^{2^{nR}}V_k(\rho^{AB})^{\otimes n}V_k^{\dagger}-\frac{{\Pi}^{A_{typ}}_{n,\epsilon}}{{\rm Tr}{\Pi}^{A_{typ}}_{n,\epsilon}}\otimes (\rho^B)^{\otimes n}\right\|_1\leq5\epsilon+8\sqrt{\epsilon}.\nonumber
\end{eqnarray}
\label{lmm:decoupling}
\end{lmm}

\begin{thm}
$K(U_{\rm Cl})=D(U_{\rm Cl})$.
\label{thm:clifcoin}
\end{thm}

\begin{IEEEproof}
Fix arbitrary $\epsilon>0$ and choose sufficiently large $n$. Let ${\bm \sigma}_{{\vec p}{\vec q}}:=\sigma_{p_1q_1}\otimes\cdots\otimes\sigma_{p_nq_n}$ be tensor products of generalized Pauli operators on ${R_B}^{\otimes n}=\bar{R}_B$. Consider an ensemble of unitaries $\{\frac{1}{d^{2n}},{\bm \sigma}_{{\vec p}{\vec q}}\}_{{\vec p}{\vec q}}$. Since $\Psi(U^{\dagger})^{R_B}=\frac{1}{d}I^{R_B}$, ${\bm \sigma}_{{\vec p}{\vec q}}$ is a unitary on the typical subspace of $(\Psi(U^{\dagger})^{\otimes n})^{\bar{R}_B}$. Because of {Schur's} lemma, the ensemble satisfies
\begin{eqnarray}
&&\frac{1}{d^{2n}}\sum_{{\vec p}{\vec q}}{\bm \sigma}_{{\vec p}{\vec q}}^{\bar{R}_B}\proj{\phi}^{\bar{R}_B}{\bm \sigma}_{{\vec p}{\vec q}}^{\dagger\bar{R}_B}=\frac{1}{d^n}I^{\bar{R}_B}.\nonumber
\end{eqnarray}
Thus, from Lemma \ref{lmm:decoupling}, if $R\geq I(AR_A:R_B)_{\Psi(U^{\dagger})}+\epsilon=K(U)+\epsilon$, there exists a set of unitaries $\{{\bm \sigma}_{{\vec p}_k{\vec q}_k}\}_{k=1}^{2^{nR}}$ on $\bar{R}_B$ such that
\begin{eqnarray}
\left\|\frac{1}{2^{nR}}\sum_{k=1}^{2^{nR}}{\bm \sigma}_{{\vec p}_k{\vec q}_k}^{\bar{R}_B}(\Psi(U^{\dagger})^{\otimes n})^{\bar{A}\bar{R}_A\bar{R}_B}{\bm \sigma}_{{\vec p}_k{\vec q}_k}^{\dagger\bar{R}_B}-(\Psi(U^{\dagger})^{\otimes n})^{\bar{A}\bar{R}_A}\otimes \frac{1}{d^n}I^{\bar{R}_B}\right\|_1\leq5\epsilon+8\sqrt{\epsilon}.\nonumber
\end{eqnarray}

When $U$ is a generalized Clifford operator, we have
\begin{eqnarray}
\sigma_{pq}^{R_B}|\Psi(U^{\dagger})\rangle^{ABR_AR_B}
&=&(U^{\dagger AB}\otimes\sigma_{pq}^{R_B})\ket{\Phi_d}^{AR_A}\ket{\Phi_d}^{BR_B}\nonumber\\
&=&e^{i\theta_{pq}}U^{\dagger AB}(I^A\otimes\sigma_{pq}^B)\ket{\Phi_d}^{AR_A}\ket{\Phi_d}^{BR_B}\nonumber\\
&=&e^{i\theta_{pq}}(\sigma_{p'q'(pq)}^A\otimes\sigma_{r's'(pq)}^B)U^{\dagger AB}\ket{\Phi_d}^{AR_A}\ket{\Phi_d}^{BR_B}\nonumber\\
&=&e^{i\theta_{pq}}(\sigma_{p'q'(pq)}^A\otimes\sigma_{r's'(pq)}^B)|\Psi(U^{\dagger})\rangle^{ABR_AR_B}.\nonumber
\end{eqnarray}
The third line follows from $\sigma_{pq}^{R_B}\ket{\Phi_d}^{BR_B}=(\sigma_{pq}^T)^{B}\ket{\Phi_d}^{BR_B}=e^{i\theta_{pq}}\sigma_{pq}^{B}\ket{\Phi_d}^{BR_B}$. In particular, we have
\begin{eqnarray}
\sigma_{pq}^{R_B}\Psi(U^{\dagger})^{AR_AR_B}\sigma_{pq}^{\dagger R_B}=\sigma_{p'q'(pq)}^A\Psi(U^{\dagger})^{AR_AR_B}\sigma_{p'q'(pq)}^{\dagger A}.\nonumber
\end{eqnarray}
{Thus, for the state $\Psi(U^{\dagger})^{AR_AR_B}$, applying $\sigma_{pq}^{R_B}$ is equivalent to applying $\sigma_{p'q'(pq)}^A$. For the same reason, for $(\Psi(U^{\dagger})^{\otimes n})^{\bar{A}\bar{R}_A\bar{R}_B}$, applying ${\bm \sigma}_{{\vec p}_k{\vec q}_k}^{\bar{R}_B}$ is equivalent to applying ${\bm \sigma}_{{\vec p}'_k{\vec q}'_k}^{\bar{A}}$.} Thus $\Psi(U^\dagger)^{AR_AR_B}$ is decoupled between $AR_A$ and $R_B$ with the randomness cost $K(U)+\epsilon$. That is, we have $D(U)\leq K(U)$. From Corollary \ref{crl:kandd}, we have $K(U_{\rm Cl})=D(U_{\rm Cl})$.
\end{IEEEproof}

\begin{figure}[t]
\begin{center}
\includegraphics[bb={0 0 729 382}, scale=0.33]{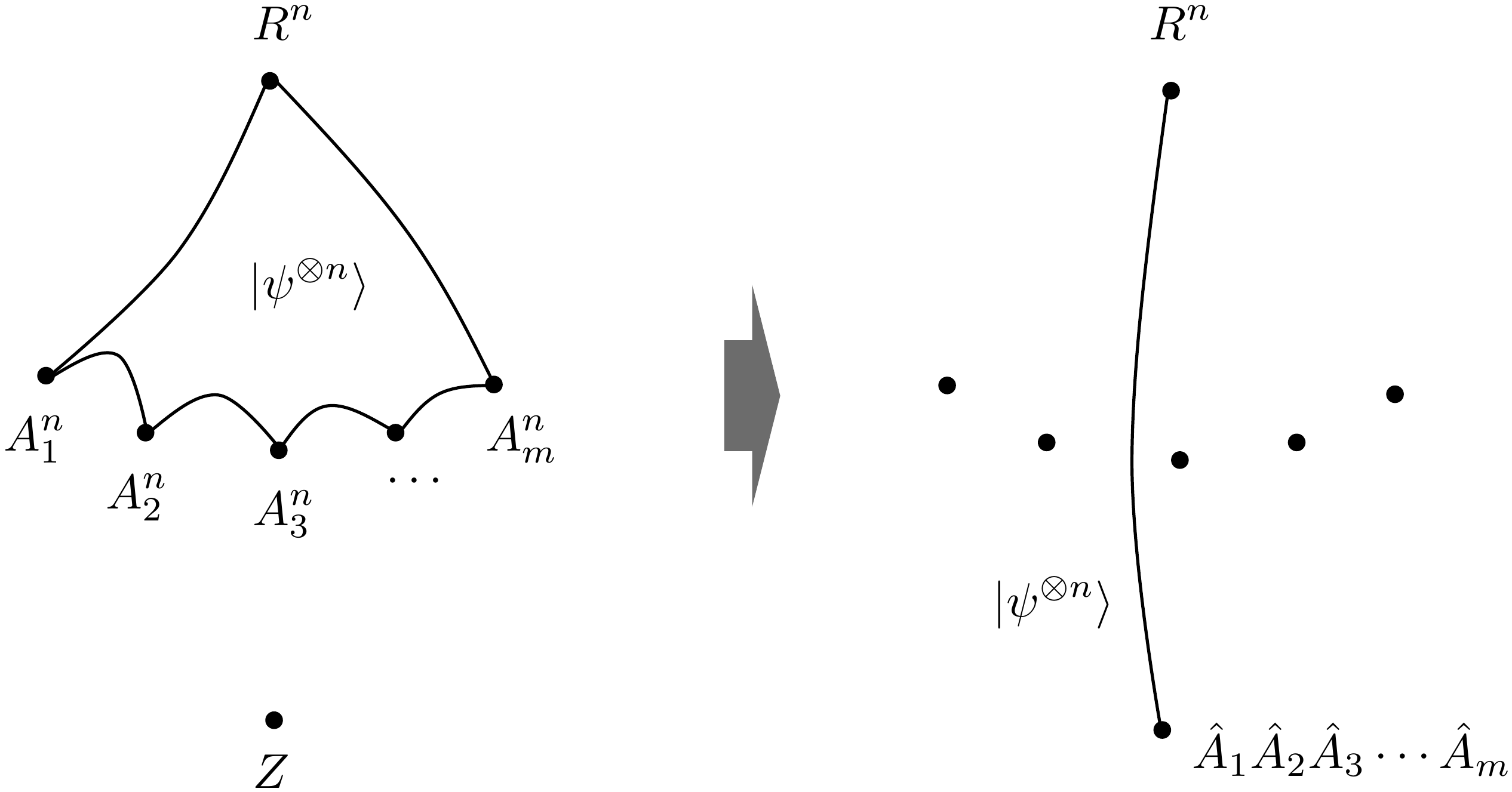}
\end{center}
\caption{In the distributed compression of $m$-partite quantum state, $m$ senders transfer their share of a quantum state to a receiver $Z$, who has quantum systems $\hat{A}_1\hat{A}_2\hat{A}_3\cdots\hat{A}_m$ isomorphic to $A_1^nA_2^nA_3^n\cdots A_m^n$. The $m$ senders are not allowed to communicate with each other.}
\label{fig:distcomp}
\end{figure}

\begin{crl}
For generalized Clifford operators $U_{\rm Cl}$, a rate triplet $(R_E,C^{\rightarrow},C^{\leftarrow})$ is achievable if and only if $R_E,C^{\rightarrow},C^{\leftarrow}\geq K(U_{\rm Cl})$.
\label{crl:clifford}
\end{crl}
{
\begin{IEEEproof}
Follows from Theorem \ref{thm:lowerbound}, Theorem \ref{thm:upperbound} and Theorem \ref{thm:clifcoin}.
\end{IEEEproof}
}

\section{Distributed Compression}
\label{sec:distributedcompression}

Distributed compression of quantum states is a task in which many parties transfer a shared state to a receiver, by using as small amount of quantum communication from each party to the receiver as possible, but with no communication among senders. Distributed compression of bipartite states is analyzed in \cite{ahn06,abey09}, and multipartite cases are formulated and considered in \cite{avis08}. In distributed compression of $m$-partite quantum state, $m$ parties $A_1,\cdots, A_m$ initially share $n$ identical copies of a state $|\psi\rangle^{A_1\cdots A_mR}$ with an inaccessible reference system $R$ (Fig. \ref{fig:distcomp}). The objective is to transfer all the senders' shares of  $(|\psi\rangle^{A_1\cdots A_mR})^{\otimes n}$ to a receiver $Z$ without destroying entanglement with the reference system. It is desirable that the total amount of quantum communication is as small as possible. We are interested in finding out the set of quantum communication rates from each senders to the receiver that are achievable under the condition that the transmission error vanishes in the limit of $n\rightarrow\infty$. 

A trivial protocol is that each sender individually performs the Schumacher compression \cite{schumacher95} on $A_1^{\otimes n},\cdots, A_m^{\otimes n}$, respectively, and send the compressed state to the receiver. This protocol apparently requires quantum communication of the rate $S(A_1)+\cdots+S(A_m)$ in total. But there are cases where they can further reduce the total quantum communication rate by exploiting correlations in the state $\psi^{A_1\cdots, A_m}$. In such cases, there are trade-off relations among the rates of quantum communication required of the senders in general. However, it is difficult to characterize multipartite correlations that are useful for reducing the communication rates. Consequently, only a few things are known on characterizing the achievable rate region in the distributed compression of multipartite quantum states.

In this section, we consider distributed compression of a class of tripartite quantum states associated with bipartite unitaries. We consider a particular situation in which one of the three senders do nothing, that is, the quantum communication rate of the sender is assumed to be zero. We prove two theorems concerning a lower and an upper bound on the quantum communication rate required in distributed compression. First, we give a lower bound in terms of the minimum entanglement cost or the minimum classical communication cost in EALOCC implementations of the unitary ({\it Theorem \ref{thm:LB}}).  Second, we derive an upper bound in terms of the partial decoupling cost $D(U)$ ({\it Theorem \ref{thm:UB}}).

Let us begin with the precise definition of distributed compression of tripartite states in the form of \cite{avis08}.

\begin{dfn}
Let $\psi^{ABC}$ be a tripartite quantum state and $|\psi\rangle^{ABCR}$ be a purification thereof. Let ${\mathcal E}_n^A:A^n\rightarrow A_S$, ${\mathcal E}_n^B:B^n\rightarrow B_S$ and ${\mathcal E}_n^C:C^n\rightarrow C_S$ be CPTP maps such that the dimension of the output systems $A_S$, $B_S$, $C_S$ are $2^{nQ_A}$,  $2^{nQ_B}$ and  $2^{nQ_C}$, respectively. Let $\hat{A}$, $\hat{B}$, $\hat{C}$ be systems that are isomorphic to $A^n$, $B^n$ and $C^n$, respectively, and ${\mathcal D}_n:A_SB_SC_S\rightarrow\hat{A}\hat{B}\hat{C}$ be a CPTP map. A set $({\mathcal E}_n^A, {\mathcal E}_n^B, {\mathcal E}_n^C, {\mathcal D}_n)$ is called a code for $(\psi^{ABC})^{\otimes n}$ with rate $(Q_A, Q_B, Q_C)$ and error $\epsilon_n$ if it satisfies 
\begin{eqnarray}
\bra{\psi^{\otimes n}}^{{\hat A}{\hat B}{\hat C}R^n}({\mathcal D}\circ({\mathcal E}^A\otimes{\mathcal E}^B\otimes{\mathcal E}^C))(\ket{\psi^{\otimes n}})\ket{\psi^{\otimes n}}^{{\hat A}{\hat B}{\hat C}R^n}\geq1-\epsilon_n.
\label{cond:distcomp}
\end{eqnarray}
\end{dfn}

\begin{dfn}
A rate triplet $(Q_A, Q_B, Q_C)$ is said to be achievable in the distributed compression of $\psi^{ABC}$ if there exists a sequence of the codes $({\mathcal E}_n^A, {\mathcal E}_n^B, {\mathcal E}_n^C, {\mathcal D}_n)$ for $(\psi^{ABC})^{\otimes n}$ with the rate $(Q_A, Q_B, Q_C)$ such that $\epsilon_n\rightarrow0$ in the limit of $n\rightarrow0$. The closure of the set of all achievable rate triplets is called the rate region. A rate triplet $(Q_A, +\infty, Q_C)$ is said to be achievable if there exists $q_b\geq0$ such that $(Q_A, q_b, Q_C)$ is achievable.
\end{dfn}

\begin{figure}[t]
\begin{center}
\includegraphics[bb={0 0 544 489}, scale=0.4]{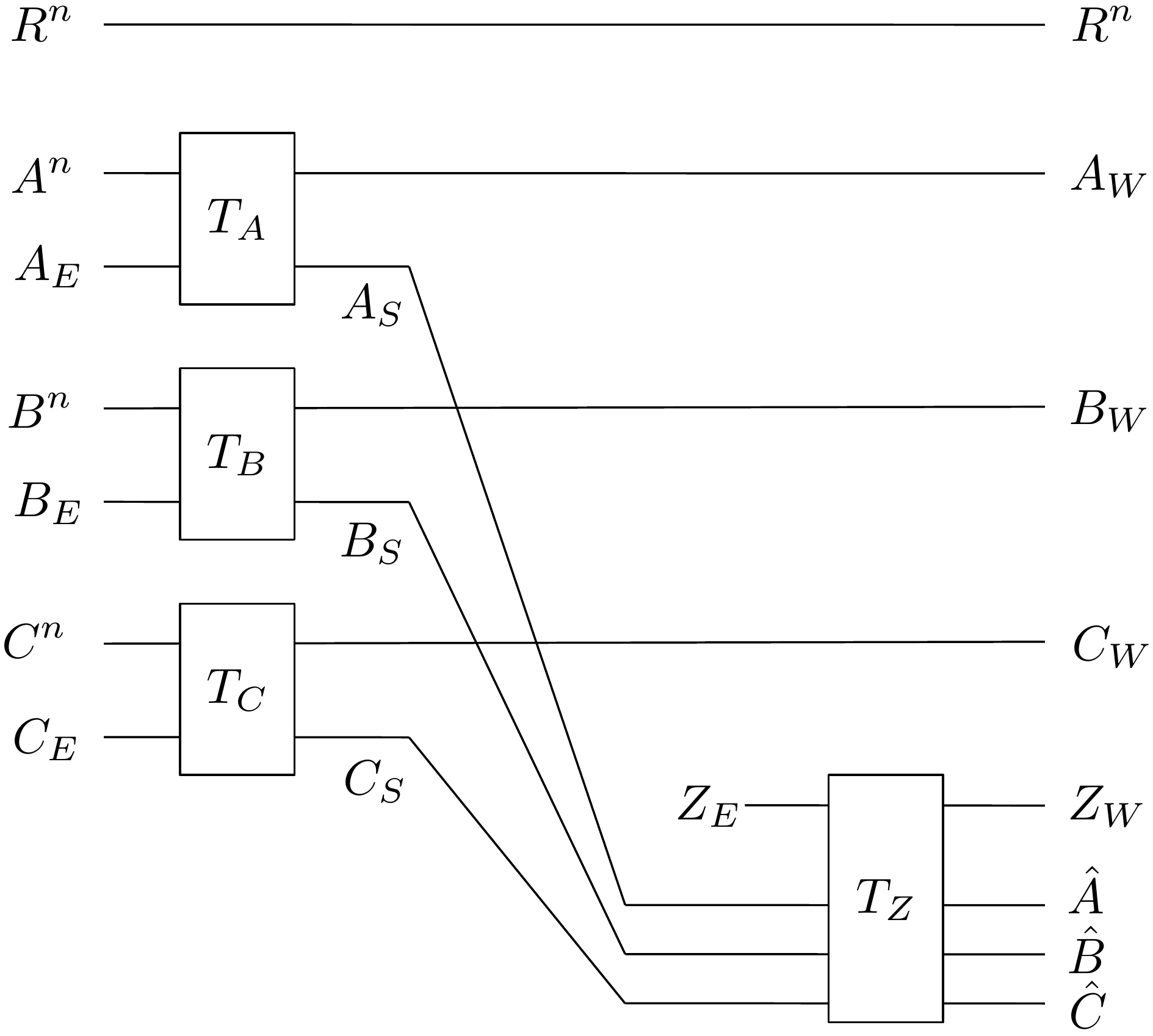}
\end{center}
\caption{The diagram describes the purified picture of the whole encoding-decoding procedure.}
\label{fig:purifydc}
\end{figure}

As is the case in the fully-quantum Slepian-Wolf protocol, it is important to consider decoupling of senders' system from the reference system \cite{avis08}. To clarify this point, consider the encoding-decoding procedure described by the encoding maps ${\mathcal E}_n^A$, ${\mathcal E}_n^B$, ${\mathcal E}_n^C$, and the decoding map ${\mathcal D}_n$ in the `purified' picture (Fig. \ref{fig:purifydc}). Let ${\rm Tr}_{A_W}[T_A(\rho^{A^n}\otimes\proj{0}^{A_E})T_A^{\dagger}]$ be the Stinespring dilation of the encoding operations ${\mathcal E}_n^A$, where $T_A$ is an isometry from $A^nA_E$ to $A_SA_W$. Similarly, let $T_B:B^nB_E\rightarrow B_SB_W$, $T_C:C^nC_E\rightarrow C_SC_W$ and $T_Z:A_SB_SC_SZ_E\rightarrow {\hat A}{\hat B}{\hat C}Z_W$ be isometries associated with the encoding operations ${\mathcal E}_n^B$, ${\mathcal E}_n^C$ and the decoding operation ${\mathcal D}_n$, respectively. After the encoding procedure, the whole purified state is given by  
\begin{eqnarray}
|\psi'_n\rangle^{A_SB_SC_SR^nA_WB_WC_W}:=(T_A\otimes T_B\otimes T_C)|\psi^{\otimes n}\rangle^{A^nB^nC^nR^n}|000\rangle^{A_EB_EC_E}.\nonumber
\label{puredistcomp}
\end{eqnarray}
It is proved in \cite{avis08} that the condition (\ref{cond:distcomp}) implies
\begin{eqnarray}
\left\|\psi_n'^{R^nA_WB_WC_W}-\psi_n'^{R^n}\otimes\psi_n'^{A_WB_WC_W}\right\|_1\leq\epsilon'_n,
\label{cond:distcomp2}
\end{eqnarray} 
where $\epsilon'_n=2\sqrt{\epsilon_n}$. Conversely, Uhlmann's theorem guarantees that if the condition (\ref{cond:distcomp2}) is satisfied for some $\epsilon'_n$, there exists a decoding operation $\mathcal{D}_n$ such that the total error is $\epsilon_n\leq2\sqrt{\epsilon'_n}$. Thus decoupling of senders' system from the reference system is a necessary and sufficient condition for the distributed compression to succeed in high fidelity.

\begin{figure}[t]
\begin{center}
\includegraphics[bb={0 0 556 206}, scale=0.44]{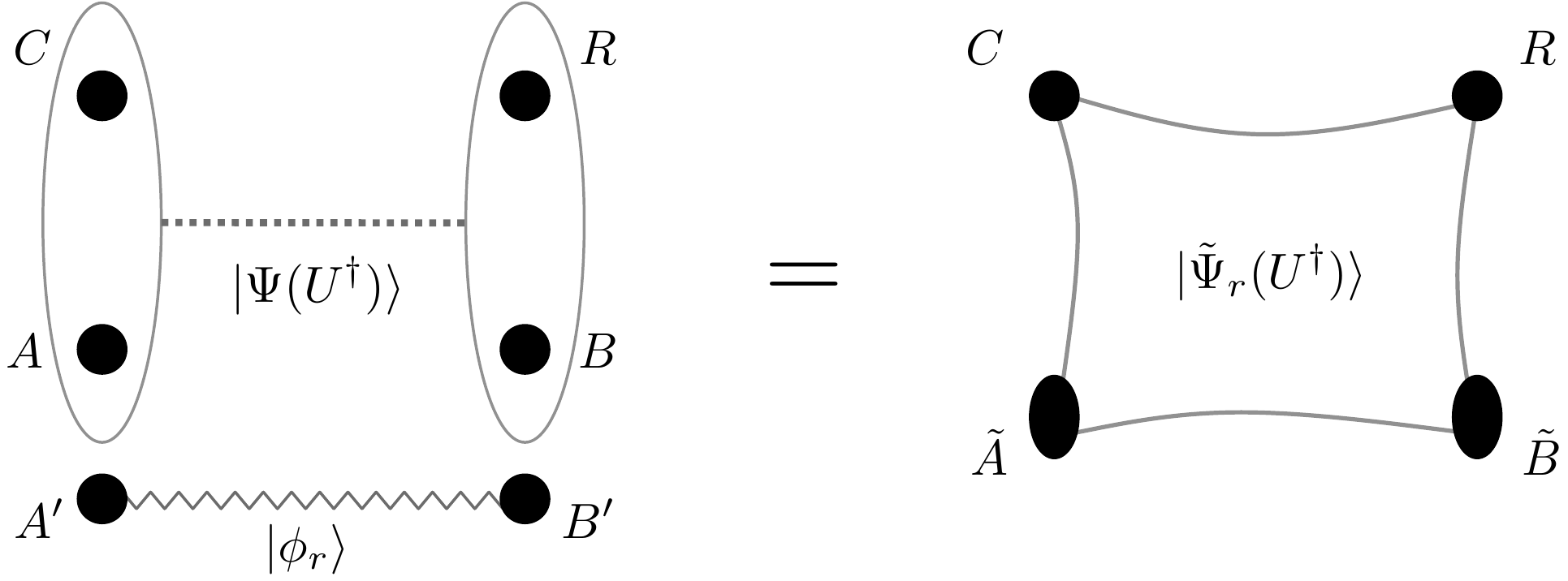}
\end{center}
\caption{The distributed compression of a tripartite state $\tilde{\Psi}_r(U^{\dagger})^{{\tilde A}{\tilde B}C}$. Its purification $|\tilde{\Psi}_r(U^{\dagger})\rangle^{{\tilde A}{\tilde B}CR}$ is composed of $|\Psi(U^{\dagger})\rangle^{{A}{B}CR}$ and $|\phi_{r}\rangle^{A'B'}$.}
\label{fig:psir}
\end{figure}

In this section, we consider distributed compression of a tripartite quantum state which is purified into $|\tilde{\Psi}_r(U^{\dagger})\rangle^{{\tilde A}{\tilde B}CR}:=|\Psi(U^{\dagger})\rangle^{{A}{B}CR}|\phi_{r}\rangle^{A'B'}$, where ${\tilde A}=AA'$, ${\tilde B}=BB'$, $|\Psi(U^{\dagger})\rangle^{{A}{B}CR}:=({U^{\dagger}}^{AB}\otimes I^{CR})|\Phi_d\rangle^{AC}|\Phi_d\rangle^{BR}$, and $\phi_{r}$ is a pure entangled state with the entanglement entropy $r$ (Fig. \ref{fig:psir}). We consider a particular situation where $Q_C=0$ and explore conditions for the rate triplet $(Q_{\tilde A}, +\infty, 0)$ to be achievable. 

\begin{thm}
\label{thm:LB}
For any $r\geq0$, the rate triplet $(Q_{\tilde A},Q_{\tilde B},Q_C)=(\frac{1}{2}R,+\infty,0)$ is achievable in the distributed compression of ${\tilde \Psi}_r(U^{\dagger})^{\tilde{A}\tilde{B}C}$ only if the rate triplet $(R_E,C^{\rightarrow},C^{\leftarrow})=(R,R,R)$ is achievable in the EALOCC implementation of $U^{AB}$. 
\end{thm}

\begin{IEEEproof}
Let $({\mathcal E}_n^A, {\mathcal E}_n^B, {\mathcal E}_n^C, {\mathcal D}_n)$ be a code for $({\tilde \Psi}_r(U^{\dagger})^{\tilde{A}\tilde{B}C})^{\otimes n}$ with the rate $(Q_{\tilde A},Q_{\tilde B},Q_C)=(\frac{1}{2}R,q_b,0)$ and the error $\epsilon_n$. $Q_C=0$ implies ${\mathcal E}_n^C={\rm id}^{C^n}$. The state after the encoding procedure, corresponding to (\ref{puredistcomp}), is given by
\begin{eqnarray}
|{\tilde \Psi}'_{r,n}(U^{\dagger})\rangle^{{\tilde A}_S{\tilde B}_SC^nR^n{\tilde A}_W{\tilde B}_W}:=(T_{\tilde A}\otimes T_{\tilde B})|\Psi(U^{\dagger})^{\otimes n}\rangle^{{\tilde A}^n{\tilde B}^nC^nR^n}|00\rangle^{{\tilde A}_E{\tilde B}_E}.\nonumber
\end{eqnarray}
Then the condition (\ref{cond:distcomp2}), by tracing out ${\tilde B}_W$, implies
\begin{eqnarray}
\left\|{\tilde \Psi}'_{r,n}(U^{\dagger})^{R^n{\tilde A}_WC_W}-\frac{1}{d^n}I^{R^n}\otimes{\tilde \Psi}'_{r,n}(U^{\dagger})^{{\tilde A}_WC_W}\right\|_1\leq\epsilon'_n.
\label{distcompdecouple}
\end{eqnarray} 
 
Construct a protocol for implementing $U^{\otimes n}$ by using the entanglement resource $|\Phi_{2^{n(R+r)}}\rangle^{A_0B_0}$ as follows. First Alice and Bob apply local operations on $A_0$ and $B_0$, and obtain $\ket{\Phi_{2^{nR}}}^{A'_0B'_0}|\phi_r^{A'B'}\rangle^{\otimes n}$. Then the whole state is
\begin{eqnarray}
\ket{\Phi_{2^{nR}}}^{A'_0B'_0}|\phi_r^{A'B'}\rangle^{\otimes n}\ket{\Psi(U^{\dagger})^{ABCR}}^{\otimes n}=\ket{\Phi_{2^{nR}}}^{A'_0B'_0}|{\tilde\Psi}_r(U^{\dagger})^{\tilde{A}\tilde{B}CR}\rangle^{\otimes n}.\nonumber
\end{eqnarray}
Alice applies the encoding operation $T_A$ on $\tilde{A}^n$, and thereby split $\tilde{A}^n$ into $A_S$ and $A_W$. Note that the dimension of $A_S$ is $2^{nR/2}$. Next she applies a controlled-generalized-Pauli gate of the form
\begin{eqnarray}
{\tilde V}^{A'_0A_S}=\sum_{k'=1}^{2^{nR}}\proj{k'}^{A'_0}\otimes\sigma_{k'}^{A_S}.\nonumber
\end{eqnarray}
Alice and Bob then perform step 2$\sim$5 in the `forward process' described in Section \ref{sec:upperbound}, which costs $nR$ bits of forward classical communication. After that, the reduced state on Alice's system and $C^nR^n$ is given by
\begin{eqnarray}
{\tilde\Psi}_{r,n}^{5}(U^{\dagger})^{{\tilde A}_S{\tilde A}_WC^nR^n}
&=&2^{-nR}\sum_{k'=1}^{2^{nR}}\sigma_{k'}^{{\tilde A}_S}\tilde{\Psi}'_{r,n}(U^{\dagger})^{{\tilde A}_WC^nR^n{\tilde A}_S}\sigma_{k'}^{\dagger {\tilde A}_S}\nonumber\\
&=&\tilde{\Psi}'_{r,n}(U^{\dagger})^{{\tilde A}_WC^nR^n}\otimes\frac{1}{2^{nR/2}}I^{{\tilde A}_S}.\nonumber
\end{eqnarray}
From (\ref{distcompdecouple}), we obtain
\begin{eqnarray}
\left\|{\tilde\Psi}_{r,n}^{5}(U^{\dagger})^{{\tilde A}_S{\tilde A}_WC^nR^n}-{\tilde\Psi}_{r,n}^{5}(U^{\dagger})^{{\tilde A}_S{\tilde A}_WC^n}\otimes\frac{1}{d^n}I^{R^n}\right\|\leq\epsilon_n.\nonumber
\end{eqnarray}
Thus Bob can obtain $(\Phi^{BR})^{\otimes n}$ in high fidelity by performing an isometry $W:\tilde{B}B_0\rightarrow B^nB_1$. 

The forward process is followed by the backward process. The entanglement cost and the classical communication cost in the latter are calculated as follows. The state of Alice's remaining systems $A_SA_W$ after the forward process is one obtained by applying an isometry $T_A$ on $\frac{1}{d^n}I^{\tilde{A}^n}\otimes\proj{0}^{A_E}$ and then performing a random generalized Pauli operation on $A_S$. Thus the entropy is $S(A_SA_W)\geq S(\tilde{A}^n)=n(r+\log{d})$. Hence the entanglement cost in the backward step is $S(B_1|A_SA_W)=S(C^n)-S(A_SA_W)\leq-nr$, which means that $nr$ ebits of entanglement is obtained after the process. On the other hand, since $B^n$ and $B_1$ are in an almost product state, we have $S(B_1)=S(B^nB_1)-S(B^n)=S(\tilde{B}B_0)-S(B^n)\leq n\log{d}+n(R+r)-n\log{d}=n(R+r)$, up to a vanishingly small error. 
Thus the classical communication cost is $I(C^n:B_1)=S(C^n)+S(B_1)-S(A_SA_W)\leq nR$. 

In total, the protocol implements $U^{\otimes n}$ with the cost of $nR$ bits of forward and backward classical communication, by using $n(R+r)$ ebits of initially shared entanglement and obtaining $nr$ ebits afterwards.
\end{IEEEproof}

\begin{thm}
The rate triplet $(Q_{\tilde A},Q_{\tilde B},Q_C)=(\frac{1}{2}R,\log{d}+\frac{3}{2}R,0)$ is achievable in distributed compression of $|\tilde{\Psi}_r(U^{\dagger})\rangle$ for any $r\geq\frac{3}{2}R$ if $R\geq D(U)$.
\label{thm:UB}
\end{thm}

\begin{IEEEproof}
Suppose $R\geq D(U)$ and consider the `forward process' to implement $U^{\otimes n}$ by using an entanglement resource $|\Phi_{2^{nR}}\rangle^{A_0B_0}$. The process requires $nR$ bits of classical communication from Alice to Bob. Since the classical message is uniformly distributed and is decoupled from the state obtained after the forward process, we can apply the {\it coherent communication identity} \cite{harrow04}. That is, we can replace $nR$ bits of classical communication by $\frac{1}{2}nR$ bits of quantum communication and an additional entanglement resource $|\Phi_{2^{nR/2}}\rangle^{A'_0B'_0}$ shared in advance. The `coherent version' of the forward process obtained in this way is composed of three steps: Alice performs a unitary on $A^nA_0A'_0$, sends $A'_0$ to Bob, and Bob performs a unitary on $B^nB_0B'_0A_0$. $A^nA_0C^n$ and $R^n$ are decoupled after this process up to a vanishingly small error.

Assume $r=\frac{3}{2}R$. In distributed compression, the senders initially share $(|\tilde{\Psi}_r(U^{\dagger})\rangle^{{\tilde A}{\tilde B}CR})^{\otimes n}=(|\Psi(U^{\dagger})\rangle^{{A}{B}CR})^{\otimes n}(|\Psi_{3R/2}\rangle^{A'B'})^{\otimes n}$. Divide  $|\Psi_{3R/2}\rangle^{\otimes n}$ into $|\Phi_{2^{nR}}\rangle^{A_0B_0}$ and $|\Phi_{2^{nR/2}}\rangle^{A'_0B'_0}$. First, Bob sends all his system $BB_0B_1$ to the receiver, which requires $n(\log{d}+\frac{3}{2}R)$ qubits of quantum communication. Next, the receiver takes over Bob's role, and Alice and the receiver performs the coherent forward process, which requires $\frac{1}{2}nR$ qubits of quantum communication. The receiver then has all the purification of the reference system, since $A^nA_0C^n$ and $R^n$ are almost decoupled. Thereby distributed compression is accomplished. Generalization to an arbitrary $r\geq\frac{3}{2}R$ is straightforward.  
\end{IEEEproof}

\begin{crl}
The rate triplet $(Q_{\tilde A},Q_{\tilde B},Q_C)=(\frac{1}{2}K(U_{\rm Cl}), \frac{3}{2}K(U_{\rm Cl})+\log{d}, 0)$ is achievable in distributed compression of $|\tilde{\Psi}_{3R/2}(U_{\rm Cl}^{\dagger})\rangle$ for any generalized Clifford operator $U_{\rm Cl}$.
\end{crl}

\begin{IEEEproof}
Follows from Theorem \ref{thm:clifcoin} and Theorem \ref{thm:UB}.
\end{IEEEproof}

{\it Example.}  Consider the two-qubit controlled-$Z$ gate $U_{\rm CZ}$, which is a Clifford gate. We have $K(U_{\rm CZ})=1$ and thus the rate triplet $(Q_{\tilde A},Q_{\tilde B},Q_C)=(\frac{1}{2}, \frac{5}{2}, 0)$ is achievable in distributed compression of $|\tilde{\Psi}_{3/2}(U_{\rm CZ}^{\dagger})\rangle$. Contrary to an argument in \cite{avis08}, it {\it is} possible to reduce $Q_{\tilde A}+Q_C$ by exploiting correlation between $\tilde{A}$ and $C$ despite the fact that there is no entanglement in $\tilde{\Psi}_{3/2}(U_{\rm CZ}^{\dagger})^{\tilde{A}C}$.

\section{Conclusion and discussion}
\label{sec:conclusion}

In this paper, we formulate and investigate compressibility of the entanglement cost and the classical communication cost in an entanglement-assited LOCC implementation of bipartite unitaries. We reveal that distributed quantum computation can be analyzed within the framework of quantum Shannon theory. In particular, we show that the decoupling approach, which is known to be a powerful tool in quantum Shannon theory, is also useful in the analysis of distributed quantum computation.

The power of bipartite unitaries as a resource for classical communication and entanglement generation is well investigated in \cite{kraus01,leifer01,bennett03,berry03,berry07,linden09,harrow10,scottc11}. Our approach is complementary to their approaches in that we address asymptotic costs for implementing unitaries rather than their power as a resource. 

The task formulated in this paper is different from either source coding, channel coding, entanglement distillation or dilution. Bipartite unitary operations can be regarded as bidirectional quantum channels. In this sense, the task is to simulate a {\it pure quantum bidirectional channel} by using pure entanglement and classical communication. It is similar yet different from quantum reverse Shannon theorem \cite{berta11, bennett13}, in which the task is to simulate a {\it noisy unidirectional quantum channel} by using pure entanglement and classical (or quantum) communication. We propose that we add {\it operations} to the framework of quantum Shannon theory, besides channels and correlations.

%



\appendices
\section{}
\label{app:avfid}
In this Appendix, we prove that the condition (\ref{eq:fidelityn}) implies (\ref{eq:ensfide}). For $n=1$, consider a quantum operation ${\mathcal E}$ on ${\mathcal S}({\mathcal H}^{A}\otimes{\mathcal H}^{B})$ defined as ${\mathcal E}(\rho)=U^{\dagger}{\hat{\mathcal M}}_1(\rho)U$. Suppose that the input to this operation is a pure state that is randomly chosen according to the Haar measure on ${\mathcal H}^{A}\otimes{\mathcal H}^{B}$. The average fidelity is defined as
\begin{eqnarray}
{\bar F}({\mathcal E}):=\int p(d\phi)F({\mathcal E}(|\phi\rangle),|\phi\rangle) .\nonumber
\end{eqnarray}
The corresponding entanglement fidelity is defined as
\begin{eqnarray}
F_e({\mathcal E}):=F(\rho({\mathcal E}),|\Phi_{d}\rangle^{AR_A}|\Phi_{d}\rangle^{BR_B}),\nonumber
\end{eqnarray}
where $\rho({\mathcal E})={\mathcal E}(|\Phi_{d}^{AR_A}\rangle|\Phi_{d}^{BR_B}\rangle)$. It is proved in \cite{schumacher96} that ${\bar F}({\mathcal E})\geq F_e({\mathcal E})$. On the other hand, we have
\begin{eqnarray}
{\bar F}({\mathcal E})=\int p(d\phi)F({\hat{\mathcal M}}_1(\phi),U|\phi\rangle)\nonumber
\end{eqnarray}
and
\begin{eqnarray}
F_e({\mathcal E})&=&F(\rho({\hat{\mathcal M}}_1), |\Psi(U)\rangle)\nonumber\\
			&\geq&F(\rho({\mathcal M}_1), |\Psi(U)\rangle|\Phi_{L_1}\rangle^{A_1B_1}).\nonumber
\end{eqnarray}
Thus we obtain (\ref{eq:ensfide}) from (\ref{eq:fidelityn}). Generalizing to an arbitrary $n$ is straightforward.

\section*{Acknowledgment}
The authors thank Akihito Soeda, Fabian Furrer, Masaki Owari, Go Kato and Tomohiro Ogawa for useful discussions.

\bibliographystyle{IEEEtran}
\bibliography{unicomp}

\end{document}